\documentclass{egpubl}

\ConferenceSubmission   

\usepackage[T1]{fontenc}
\usepackage{dfadobe}  

\usepackage{cite}  
\BibtexOrBiblatex
\electronicVersion
\PrintedOrElectronic
\ifpdf \usepackage[pdftex]{graphicx} \pdfcompresslevel=9
\else \usepackage[dvips]{graphicx} \fi

\usepackage{egweblnk}
\usepackage{amsmath}
\usepackage{xcolor}
\usepackage{listings}
\usepackage{graphicx, subfig}
\usepackage{wrapfig}

\title{Wavelet Transparency}
\author[Anonymous]{}
\author[Maksim Aizenshtein  \& Niklas Smal \& Morgan McGuire]
{\parbox{\textwidth}{\centering Maksim Aizenshtein ~~~~~~ Niklas Smal  ~~~~~~ Morgan McGuire\\
~~~NVIDIA ~~~~~~~~~~  UL Benchmarks~~~~~~~~~~~~~~ Roblox}\\
{\parbox{\textwidth}{\centering December 1, 2021}}}

\begin{document}
    

\maketitle

\begin{abstract}
Order-independent transparency schemes rely on low-order approximations of transmittance as a function of depth. We introduce a new wavelet representation of this function and an algorithm for building and evaluating it efficiently on a GPU. We then extend the order-independent Phenomenological Transparency algorithm to our representation and introduce a new phenomenological approximation of chromatic aberration under refraction. This generates comparable image quality to reference A-buffering for challenging cases such as smoke coverage, more realistic refraction, and comparable or better performance and bandwidth to the state-of-the-art Moment transparency with a simpler implementation. We provide GLSL source code.
\end{abstract}

\section{Introduction and Related Work}

``Transparency'' in computer graphics broadly covers all situations in which multiple surface layers or a continuous medium contribute to the value of a pixel. This includes otherwise-opaque surfaces partly covering a pixel (a.k.a., edge antialiasing, alpha masking), the gradual obscuring of objects due to smoke volumes and particles (a.k.a., partial coverage/alpha), the refraction and modulation of a background by glass, and the diffusion of light passing through fog and ground glass, as well as shadows cast in such scenes.

Real-time rendering largely relies on provisioning parallel resources proportional to the number of pixels. This works well for opaque triangles that are larger than the pixels. Real-time transparency is challenging because it is a case where screen size is not a bound on the amount of processing required, since many primitives may affect each pixel. Fortunately, human observers appear to be extremely tolerant of error in transparency in general situations, and especially in entertainment applications. This creates significant room for approximation, which has been exploited by decades of clever real-time transparency solutions and centuries of fine art. 

\vspace{-2mm}
\subsection{Ray Tracing}
The ultimate solution for transparency is path tracing, as has been demonstrated by the quality of offline rendering and relative elegance of its implementation~\cite{Pharr2016PBR}. We expect that algorithmic and hardware improvements some day will enable real-time path tracing for complicated transparent objects. Today's fastest GPUs can trace and shade 1-10 million rays per millisecond, which enables ray-traced reflections and shadows in currently available video games. However, path tracing and even Whitted ray tracing currently are too expensive for complex transparent materials, which require many more rays per pixel than opaque materials.

For example, consider a simple bottle of wine in an otherwise empty room. At least five ray intersections per pixel are required to render this scene: air to glass, glass to wine, wine to glass, glass to air, and then air to the wall behind the bottle. If each of those intersections spawns recursive shadow and reflection rays, then at least 15 rays per pixel are required and the ray cast and shade cost  quickly consumes the entire frame budget at 1080p 60 Hz. Furthermore, this is the simplest case of path tracing for perfect specular reflection and refraction. For the more difficult case of rough, diffusing surfaces such as ground glass and participating media such as fog, path tracing requires many stochastic samples to converge. For opaque surfaces, it is common to take fewer samples, which leave noise, and then apply spatio-temporal denoising methods (e.g., \cite{Mara:2017:EDA:3105762.3105774,Schied:2017:SVF:3105762.3105770}). However, those denoisers do not work well for transparent materials because the noise is at varying depths and temporal reprojection is impractical. 

\vspace{-2mm}
\subsection{The Visibility Function}
\label{sec:visbilitydefinition}
The transparency \textit{visibility function} maps distance along a view ray to net transmittance from that distance back to the ray origin. For example, in a scene with a single, 25\% opaque plane 1m from the camera, the visibility function is $v(x)=1$ on $x=[0\mathrm{m}, 1\mathrm{m})$ and then $v(x)=0.75$ for $x>1\mathrm{m}$ (Figure~\ref{fig:graphwinebottle}). Opaque rendering is the subcase in which the visibility function is binary. This compositing notion of transparency of grew out of Deep Shadow Maps~\cite{Lokovic:2000:DSM:344779.344958} in which visibility is relative to a light source, but is now applied to all visibility problems. The visibility function also can be computed separately for discrete light frequencies to model colored transmission and absorption, and the ``ray'' can be bent into a path or  extended into a cone to model specular and diffuse refraction.

\subsection{OIT Methods}
\vspace{-0.5mm}
In contrast to earlier approaches (see the survey in \cite{RTR4}), the \textit{order-independent transparency} (OIT) methods preferred for real-time transparency today do not require storing or sorting all surfaces at a pixel. This enables constant space per pixel and linear time in the number of pixels covered collectively by the primitives. Grouped roughly by their representation of the visibility function, the main OIT methods include:

\begin{description}
\item[Depth peeling] Exact, by exhaustive evaluation~\cite{Everitt01interactiveorder-independent}
\item[Stochastic] Binary visibility, with depth chosen by importance sampling \cite{SA09,Enderton2010Stochastic,McGuire11CSSM}
\item[Occupancy map] Samples at uniform depths \cite{Kim:2001:OSM:647653.732282,Sintorn:2009:HSS:1507149.1507160}
\item[$k$-buffer]  Samples at $k$ arbitrary depths, with varying strategies for choosing and merging samples \cite{Jouppi:1999:ZEH:311534.311582,Bavoil07kbuffer,Vasilakis2014KBuffer,Maule:2013:HT:2448196.2448212,Salvi2011,Salvi:2014:MAB:2556700.2556705,Crassin:2015:AGA:2699276.2699285,Crassin:2018:CSV:3231578.3231584}
\item[Fourier opacity map] Low-order Fourier basis \cite{Jansen2010}
\item[Weighted, blended OIT (WBOIT)] Constant-absorption exponential or polynomial \cite{Meshkin07,Bavoil08orderindependent,McGuire2013Transparency}
\item[Moment] Statistical moments \cite{Donnelly2006VSM,Peters2015Shadow,Peters2017MSM,Sharpe2018Moment} 
\end{description}

These all can be extended to perform not just partial coverage but also correct modulation of color during transmission by processing color channels coverage independently; efficient solutions have been demonstrated for stochastic~\cite{McGuire11CSSM} and WBOIT \cite{McGuire2013Transparency}.

Although developed for rasterization systems, these OIT methods even allow limited ray tracing of partial coverage transparency along a straight line. For example, \textit{Battlefield V} uses WBOIT to accumulate shading along particles using an ``any hit'' ray cast, rather than performing a true ``closest hit'' cast or stochastic path tracing which, as discussed in the introduction, would be prohibitively expensive today~\cite{Battlefield}.

\subsection{Refraction}

There are several methods for approximating refraction in real-time rendering, which the OIT methods from the previous section do not address \cite{Wyman2005,Sousa2005Refraction,Rousiers2012Refraction,Ganestam2015Refraction}. 

\begin{wrapfigure}{r}{32mm}
\vspace{-3.5mm}
\fbox{\includegraphics[width=1.2in]{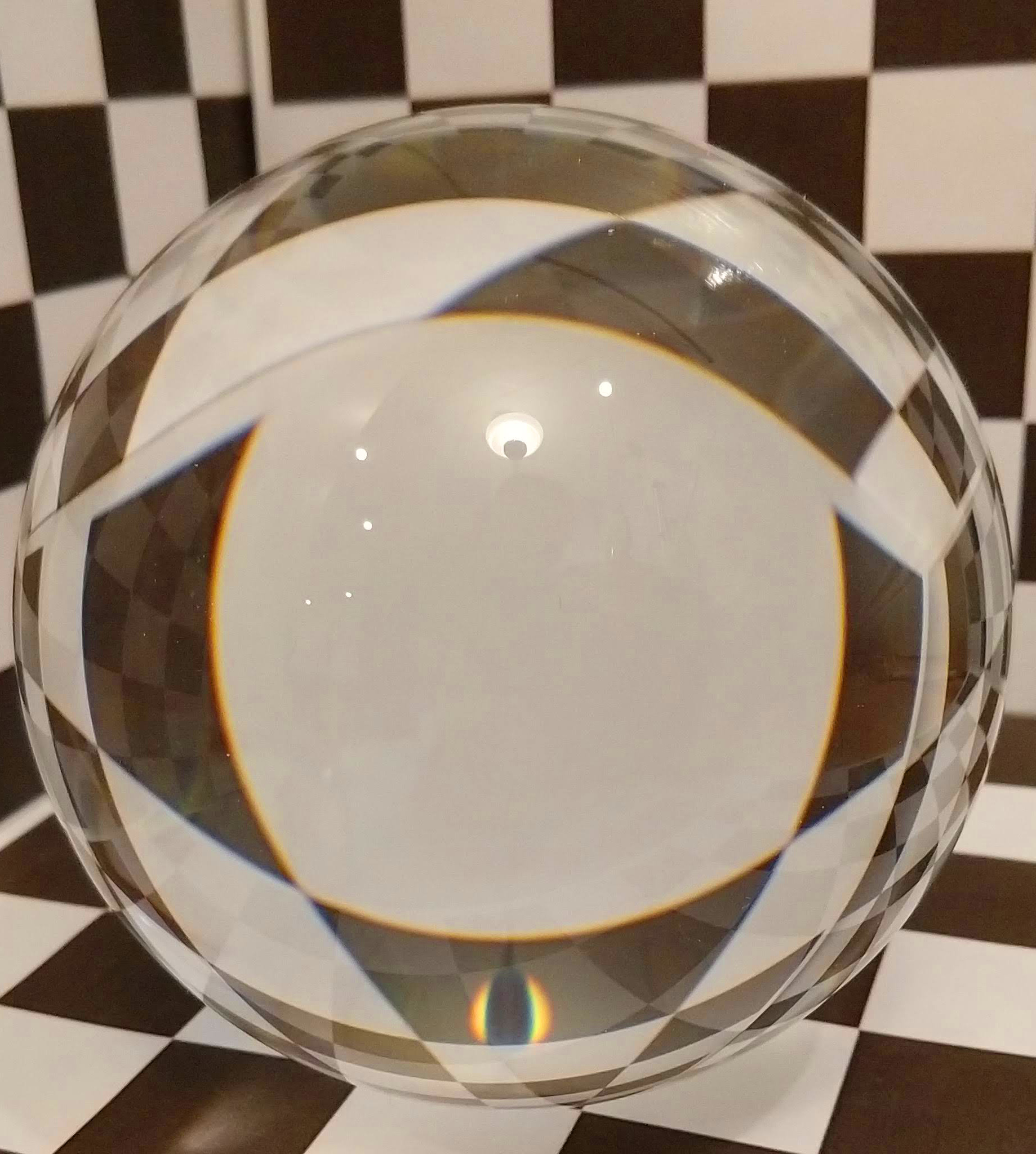}}
\vspace{1mm}
\caption{A glass ball.}
\label{fig:aberration}
\vspace{-4mm}
\end{wrapfigure}

Chromatic aberration occurs due to the index of refraction varying with the frequency of light. This produces the phenomena of colored fringes at edges in strong refractions (photographed in Figure~\ref{fig:aberration}), rainbows, and the spectral spread observed with a prism. It is trivial to implement in a ray tracer that samples light frequencies separately, and very approximate chromatic aberration due to the camera lens is a common post-processed effect in video games~\cite{Gotanda:2015:RRP:2776880.2792715}.

\vspace{-2mm}
\subsection{Other Phenomena}
\vspace{-1mm}
There are many targeted solutions for transparency in specific contexts, for example, fog \cite{Nishita1993Earth,Nishita1998Scattering,Narasimhan2003Weather,Narasimhan04analyticrendering,Premoze2004Multiple,Hillaire2015}, caustics \cite{Wyman2006,ShahKP07,Wyman2008I3D,Wyman2008JGT,Yuksel2009}, skin \cite{Jimenez2010Subsurface}, and hair \cite{Sintorn:2009:HSS:1507149.1507160,Yuksel2010HairCourse}. These outperform general transparency methods for their specific case.

Phenomenological Transparency~\cite{McGuire2017Transparency} attempts to present a unified solution for real-time transparency by combining weighted, blended OIT and colored stochastic shadow maps~\cite{McGuire11CSSM} with new order-independent approximations for refraction, diffusion, ambient occlusion, shadowing, and caustics. It gains efficiency from approximating specific phenomena important to perception of transparency rather than performing physically-accurate simulation of all light rays. 

\vspace{-2mm}
\subsection{New Contributions}
We extend Phenomenological Transparency with (1) an improved visibility representation and (2) a minor phenomenological approximation of chromatic aberration at transparent surfaces. 

The new visibility representation uses wavelets~\cite{Mallat2008Wavelets}, a hierarchical basis transformation combining the compressibility of frequency methods with the locality of spatial methods. This suits the visibility function, which exhibits sudden local changes for surfaces such as glass as well as continuous variation for materials such as fog. As with Fourier Opacity Maps, Deep Shadow Maps, and Moment Transparency and unlike other OIT methods, our representation requires separate passes over geometry to build the visibility approximation and then perform shading using it. We suggest some ways to eliminate this limitation as future work.

The new chromatic aberration term adjusts the gather samples for each color channel based on the strength of refraction and heuristics for identifying the ``front'' of objects. This is similar to previous ad hoc methods observed in video games for specific objects. It is the first addition of chromatic aberration into a general real-time OIT method that we are aware of in the literature.

Our results section shows that these new ideas combine with the previous strengths of Phenomenological and Moment Transparency to render more robustly than alternative OIT algorithms while also creating improved perception of transparency phenomena in many cases. The performance of this new Wavelet OIT depends on a design that caters to human perception rather than radiometric error, so it is not appropriate for predictive rendering and we make no claim of a direct relationship or objective comparison to physically-based simulation. Instead, we evaluate the method based on objective space and time efficiency and present visual comparison to real-time alternatives for the reader to consider for their application.

\section{Wavelet Algorithm}

\subsection{Overview}
The wavelet model of compositing has three steps, similar to MLAB and Moment Transparency. The first step is two unordered passes over all transparent geometry or their bounding boxes to build the visibility approximation, which is stored as wavelet coefficients. The second step is another unordered pass over all transparent geometry to shade it and weigh it by the visibility function. The third step composites the accumulated transparent results over the opaque parts of the scene, which is where the other effects from Phenomenological Transparency are applied.

%


\clearpage
The algorithm is:
\begin{enumerate}
    \item Compute tight per-pixel depth bounds by drawing transparents (or bounds) into a depth buffer.
    \item Construct the wavelet coefficient buffer for transparents.
    \item Accumulate transparents by drawing modulated by the visibilty function, as with previous methods.
    \item Composite over opaques and apply refraction, diffusion, etc.
\end{enumerate}

For wavelets of rank $N$ we store $2^{N+1}$ coefficients per channel, packed into 32 bits with a shared exponent in the form \texttt{E5B9G9R9}. See our code supplement for the bitpacking implementation. For example, rank $N=3$ wavelets encode 48 coefficients in 64 bytes per pixel total.

\subsection{Derivation}
The radiance compositing equation is
\[
    L = \sum_i L_i r_i\, v\left( x_i\right),
\]
where $L_i$ is the incoming radiance at $i$th interface,
$r_i$ is the reflectance, and $v\left( x_i\right)$ is the net transmittance visibility function
at depth $x_i$. Because transmittance decays exponentially with successive surfaces, 
for numerical stability convert to the net absorbance function:
\[
    A(x) = - \log{v(x)}
\]

For discrete interfaces such as glass, the absorbance function is:
\[
    A = \sum_i \alpha_i \theta\left( x - x_i\right) 
\]
where $\theta\left( x\right)$ is Heaviside theta function, and it is smooth for continuous media such as fog.
We observe that Haar wavelets are a good representation of $\theta$.

We now derive a wavelet transformation for the visibility function, and then provide an implementation and results. The code extends previous work from \cite{McGuire2017Transparency} and \cite{G3D17}.

\subsection{Wavelet approximation}
The purpose of this section is to approximate $A$ with a linear interpolation of Haar wavelets.
First we start with a direct approximation of $A$ with Haar wavelets.
\[
    A \approx \chi_{0,0} \phi_{0,0}\left( z\right) + \sum_{n,k} X_{n,k} \psi_{n,k}\left( z\right)
\]

The approach will be based on two algorithms. One for efficient evaluation of the coefficients, and another one
for efficient evaluation of the approximation at a point.

\subsubsection{Coefficients}

The coefficients are evaluated directly with integration by parts:
\begin{gather*}
    \chi_{0,0} = \int_0^1 dx\; A\cdot \phi_{0,0} =
    \\ = \left[ A\cdot \Phi_{0,0}\right]^1_0 - \int_0^1 dx\; \frac{dA}{dx}\cdot\Phi_{0,0} =
    \\ = \sum \alpha_i - \sum \int_0^1 dx\; \alpha_i \delta\left( x-x_i\right) \cdot x =
    \\ = \sum \alpha_i \left(  1 - x_i\right)
\end{gather*}

\begin{gather*}
    X_{n,k} = \int_0^1 dx\; A\cdot \psi_{n,k} =
    \\ = \left[ A\cdot \Psi_{n,k}\right]^1_0 - \int_0^1 dx\; \frac{dA}{dx} \cdot \Psi_{n,k} =
    \\ = - \sum \alpha_i \Psi_{n,k}\left( x_i\right)
\end{gather*}

and:
\[
\Psi_{n,k}\left( x\right) = 2^{-\frac{n}{2}} \cdot \left\lbrace
\begin{matrix}
2^n x - k & 0 \leq 2^n x - k < 0.5\\
1 + k - 2^n x & 0.5 \leq 2^n x - k \leq 1\\
0 & \textrm{Else}
\end{matrix}
\right.
\]

The functions $\Psi_{n,k}$ are non negative on the interval $\left[ 0,1\right]$, and are localized as their
corresponding wavelets. Overall obtaining that all wavelet coefficients are non positive, while the scaling
function coefficient is non negative.

So if the approximation is done only with wavelets and a single scaling function, then a logarithmic
number of coefficients needs to be updated per incoming interface. Same applies during evaluation of the function.

The choice of the nature of the depth has a direct effect on the wavelet basis that should be used. For instance
linear depth will require uniform wavelet scale, while screen space depth will work better
with increased resolution towards where the information is compressed. In this work the approach that is chosen
is the one with linear depth.

\subsubsection{Interpolation}
Direct approximation with Haar wavelets will produce a staircasing which will not coincide with the steps
of the original function. To alleviate this issue, but still keep a relatively smooth and predictable approximation,
the approximated function can be linearly interpolated between the points on the subdivision.

\subsubsection{Coefficient update}
Simplest implementation that will preserve coherency can be based on ROVs, however in this case atomic operations
can be used as well. If the hardware supports floating point atomic operations, then the implementation is trivial.
If not, then a CAS loop can be done per coefficient.

\section{Chromatic Aberration Algorithm}
Phenomenological Transparency additively accumulates a screen-space refraction offset vector at each pixel during 
step 2 (shading) of transparency rendering. This vector
can be derived from ray cast against a fixed background plane, a background plane at the opaque depth of the current pixel
(used in this paper's results), or via true geometric or screen-space ray casting against the full depth buffer. 
During the step 3 (compositing), Phenomenological Transparency uses this offset vector for the texture sample of the background image of opaque surfaces. Our new chromatic aberration approximation affects both steps. 

During the shading step, it cubes the transmission coefficient when the refractive index is higher than 1. This is a heuristic for
crudely approximating an extinction coefficient within the medium in a order-independent way. It causes the back surface of
a transparent object surrounded by a lower-index medium (e.g., a glass in air or water) to be darker, as if there were absorption
while the ray passed through the object. In particular, this causes reflections on the inner surface of an object to be slightly 
darker than on the outer surface, as observed in real scenes. The inverted reflection of the lamp on the lower part of the 
real ball in Figure~\ref{fig:aberration} is an example of this phenomenon; it is dimmer than the right-side-up specular highlight
on the top part of that ball.

During the compositing step, the algorithm replaces the single bilinear sample of the background image with $k=5$ bilinear samples along a short line segment, aligned with the refraction vector and with proportional magnitude. For each tap index $0 \leq i < k$, we weight the color channels by:
\begin{verbatim}
vec3 spectralWeight(int i) {
    float t = 0.5 + float(2 * i) / float(k - 1);
    vec3 w;
    w.r = smoothstep(0.5, 1./3, t);
    w.b = smoothstep(0.5, 2./3., t);
    w.g = 1.0 - w.r - w.b;
    return w;
}
\end{verbatim}

\begin{figure}[tb!]
\begin{center}
\subfloat[A car in fog]{\includegraphics[width=0.95\columnwidth,clip,trim=4mm 9cm 14cm 3cm]{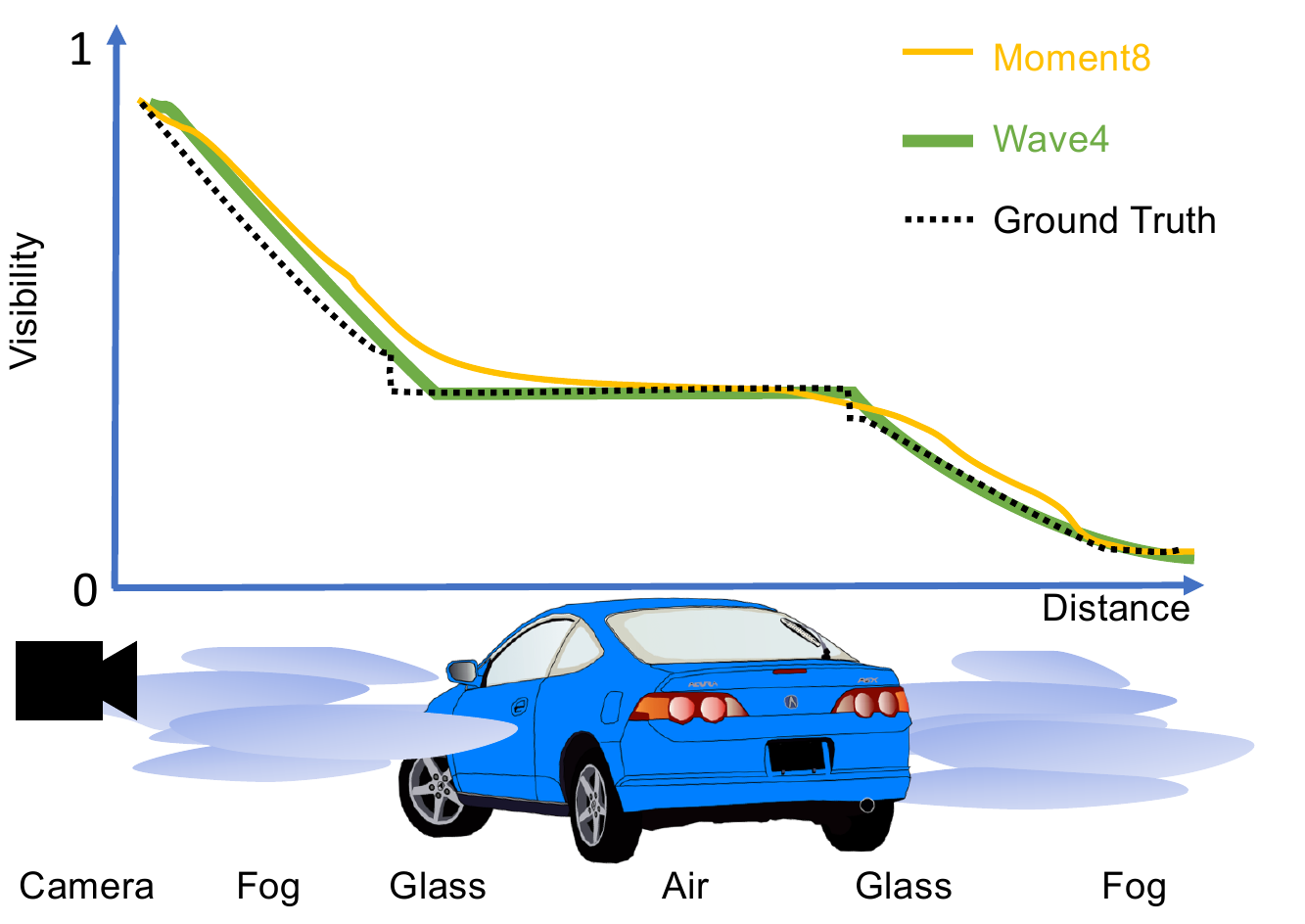}
\vspace{-1mm}
\label{fig:graphcarfog}}
\vspace{-5mm}

\subfloat[A glass bottle of wine]{\includegraphics[width=0.95\columnwidth,clip,trim=4mm 9cm 14cm 3cm]{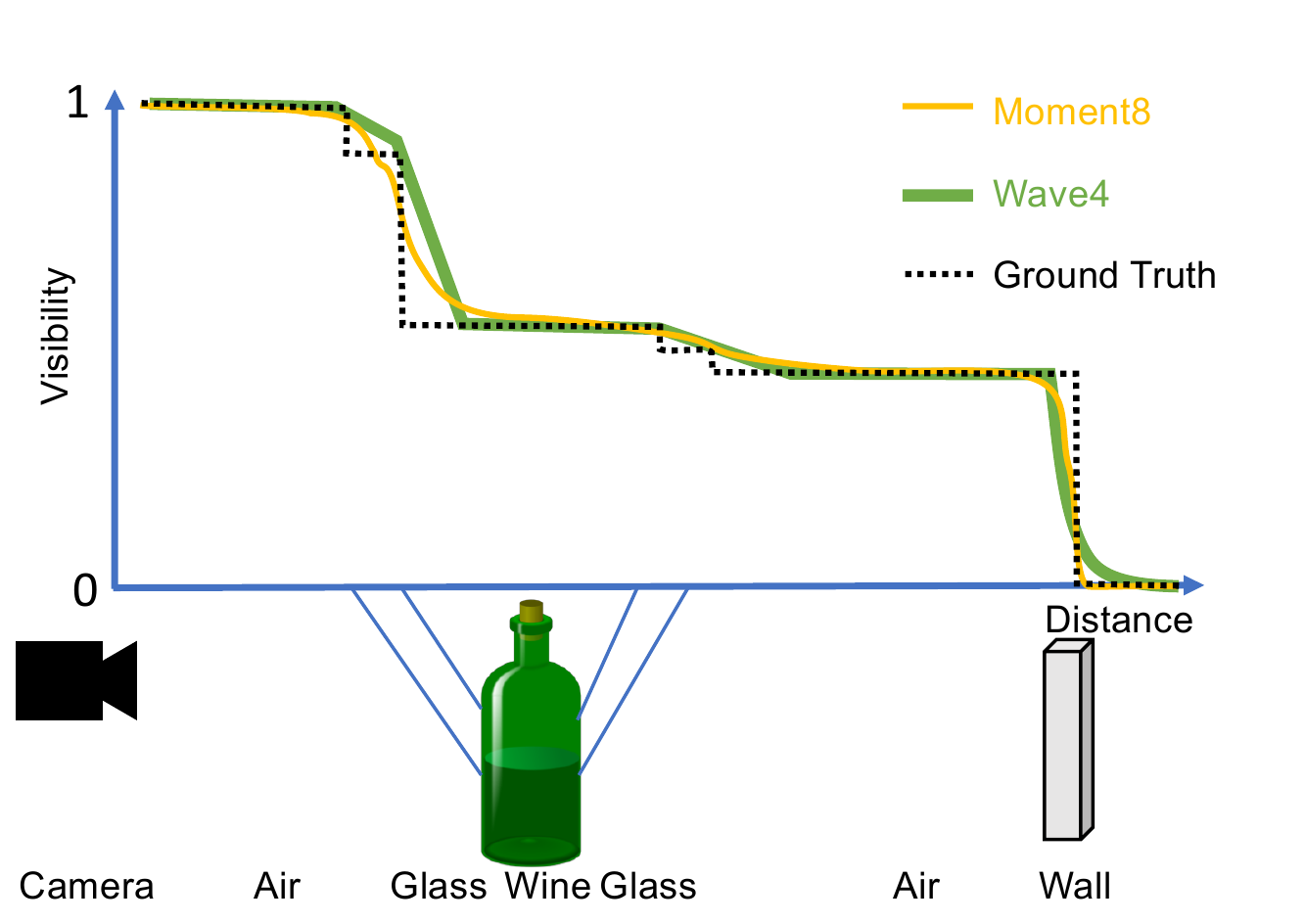}
\vspace{-1mm}
\label{fig:graphwinebottle}}
\vspace{-5mm}

\subfloat[Two particle systems]{\includegraphics[width=0.95\columnwidth,clip,trim=4mm 9cm 14cm 3cm]{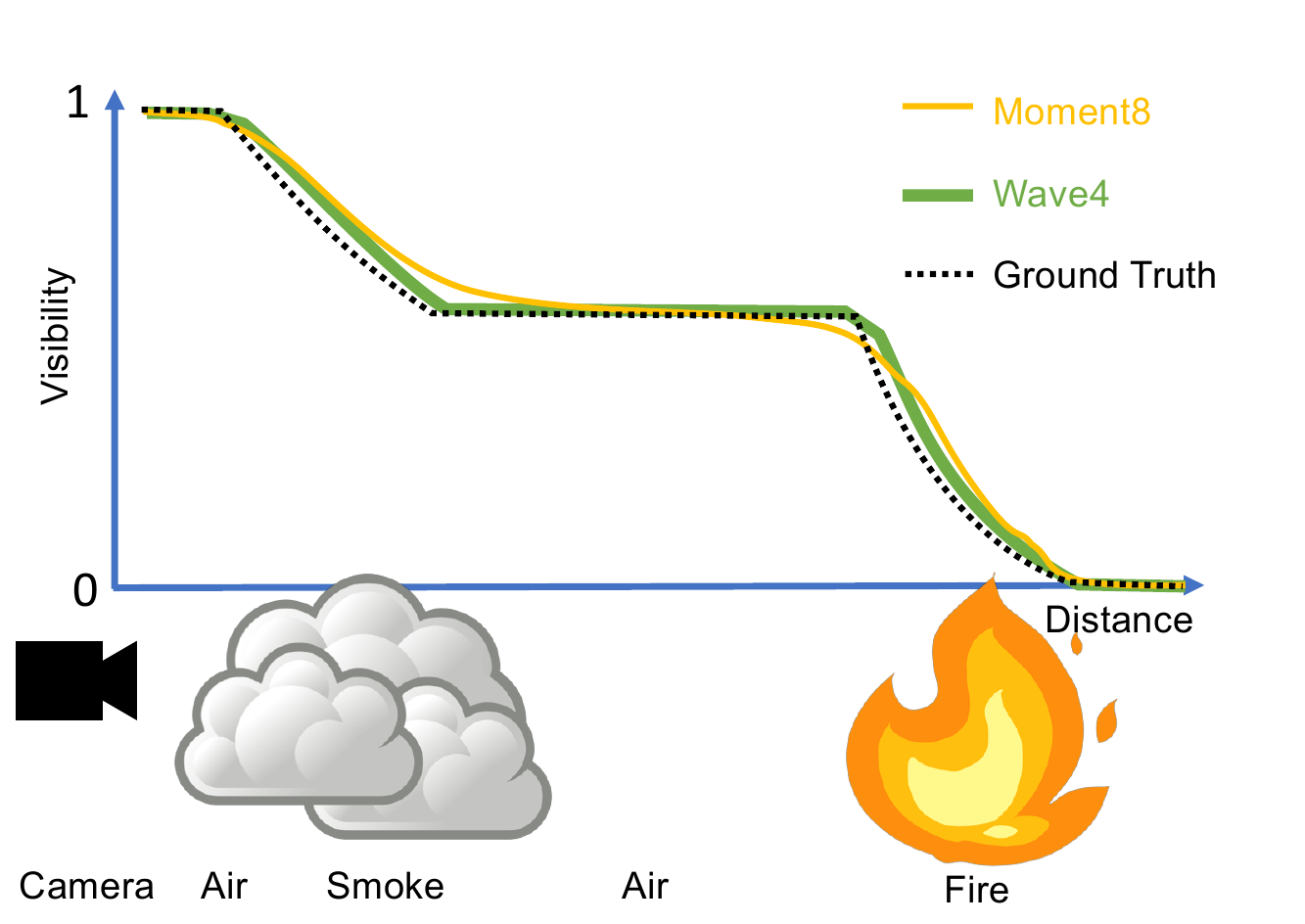}
\vspace{-1mm}
\label{fig:graphsmokeandfire}}

\caption{Comparison of visibility function approximations.}
\label{fig:visibility}
\end{center}
\vspace{-5mm}
\end{figure}

\section{Results}

\subsection{Visualizing Visibility}
\label{sec:visibility}

To understand the nature of the approximations, we graph the visibility function along the central ray for some everyday scenes with transparency in Figure~\ref{fig:visibility}. The figures in this section display the scenes as cartoon schematics and the graphs as computed from the equivalent real 3D scene.

In the figure, ground truth visibility is a dashed black line, Moment Transparency is yellow, and Wavelet Transparency is a thick green line. We compare at parameters that yield approximately equal performance for them.

Figure~\ref{fig:graphcarfog} shows the view along a ray through the windows of a car that is surrounded by fog. The ground truth visibility function decreases in a shallow exponential within fog, is constant in air, and suddenly drops at the surface of the glass windows. Although both approximations slightly overestimate visibility within fog near the camera, Wavelet is a strictly better approximation than Moment across the entire scene.
Figure~\ref{fig:graphwinebottle} shows visibility in the bottle of winebottle scene described in Section~\ref{sec:visbilitydefinition}, modeled without extinction within the wine. Here, both approximations oversmooth and Moment gives a tighter fit near the first interface.
Figure~\ref{fig:graphsmokeandfire} shows separate smoke and fire particle systems. Wavelet is a better fit overall and captures the sharp transitions slightly better.

\setlength{\tabcolsep}{2pt}
\begin{figure*}[tb]
\centering
\begin{tabular}{lcccc|c}
& MLAB4~\protect{\cite{Salvi:2014:MAB:2556700.2556705}} & WBOIT~\protect{\cite{McGuire2013Transparency}} & Moment~\protect{\cite{Sharpe2018Moment}}  &  Wavelet~(New) & A-Buffer~\protect{\cite{Carpenter1984ABuffer}}\\
\!\!\!1\!\!\! &
\fbox{\includegraphics[width=0.19\textwidth,clip,trim=0 1cm 0 2.5cm]{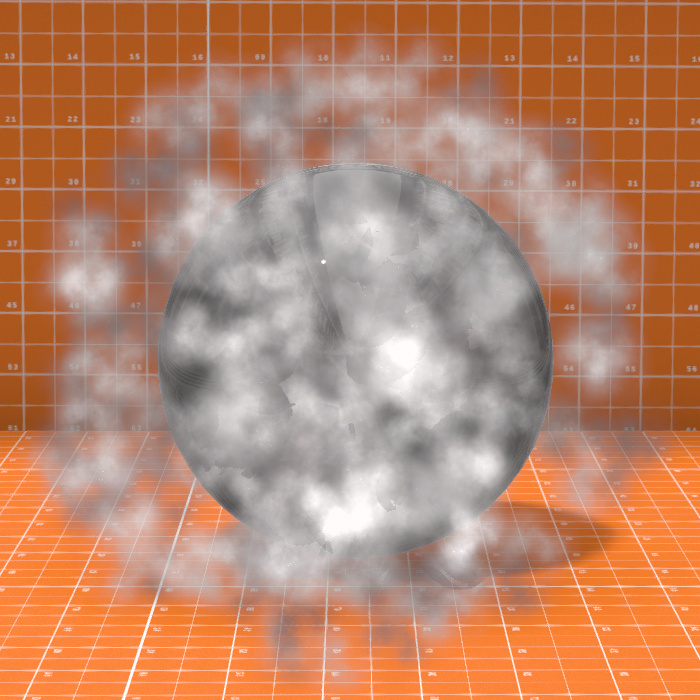}} &
\fbox{\includegraphics[width=0.19\textwidth,clip,trim=0 1cm 0 2.5cm]{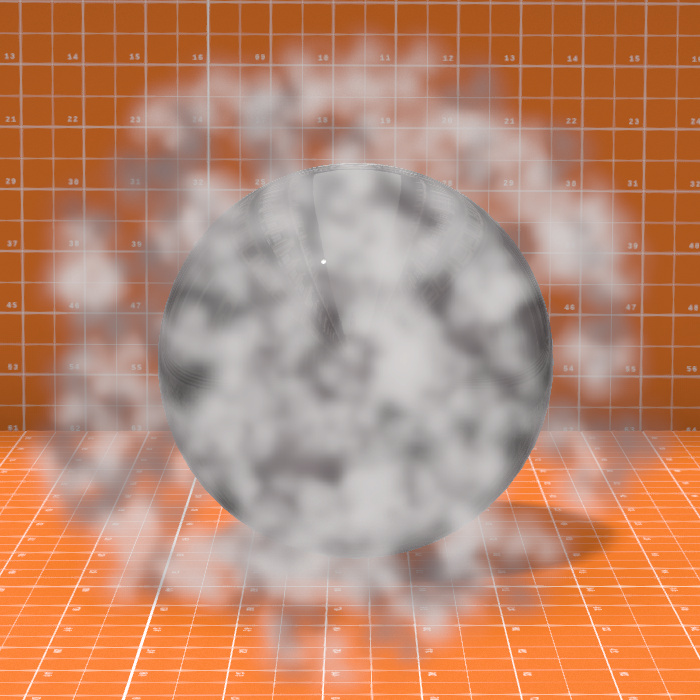}} &
\fbox{\includegraphics[width=0.19\textwidth,clip,trim=0 1cm 0 2.5cm]{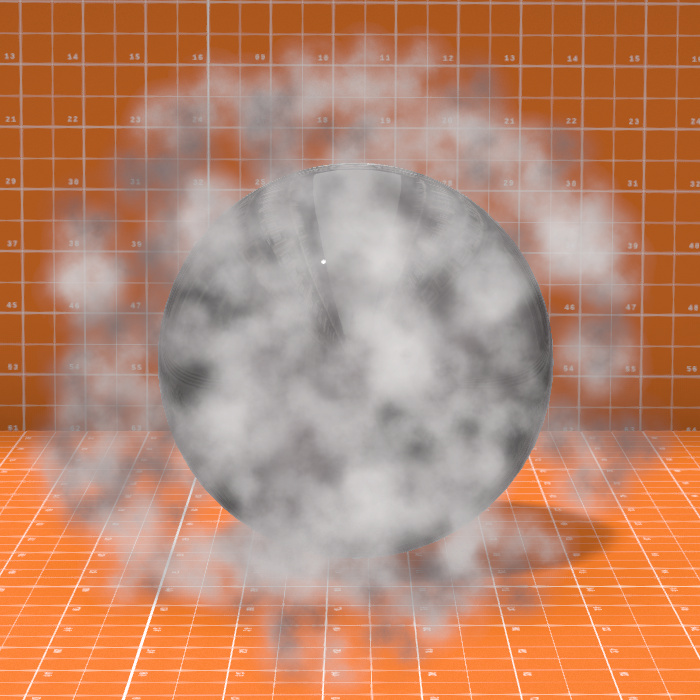}} &
\fbox{\includegraphics[width=0.19\textwidth,clip,trim=0 1cm 0 2.5cm]{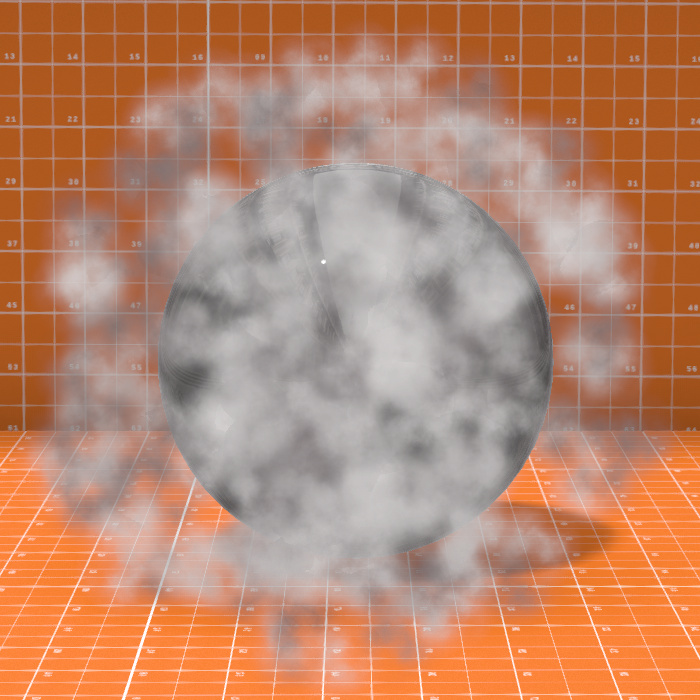}} &
\fbox{\includegraphics[width=0.19\textwidth,clip,trim=0 1cm 0 2.5cm]{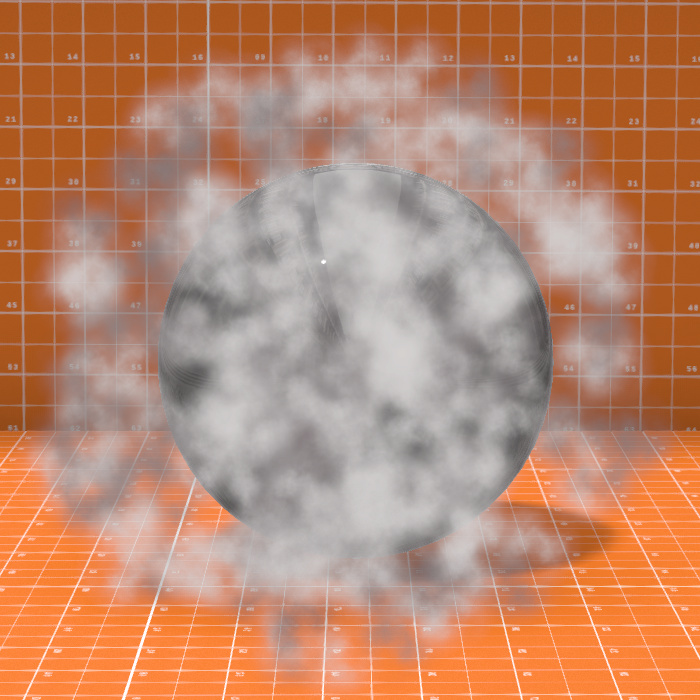}} \\

\!\!\!2\!\!\! &
\fbox{\includegraphics[width=0.19\textwidth,clip,trim=0 1cm 0 2.5cm]{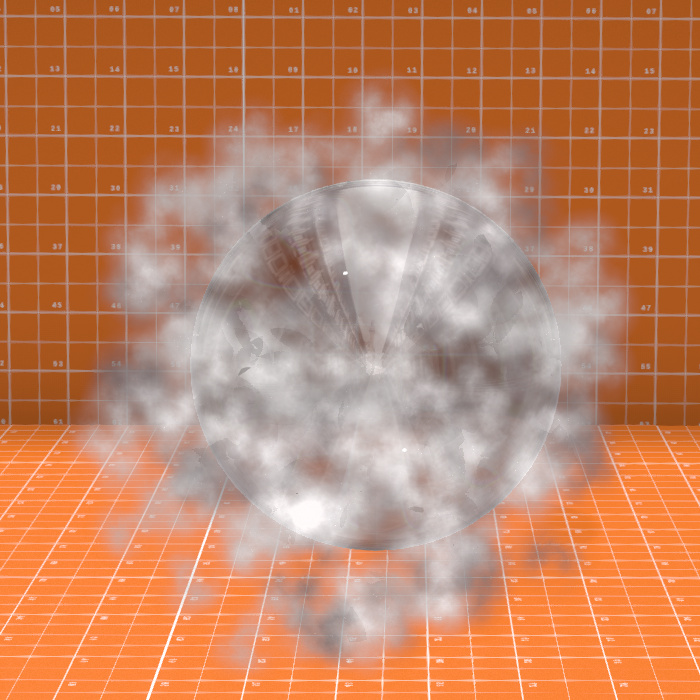}} &
\fbox{\includegraphics[width=0.19\textwidth,clip,trim=0 1cm 0 2.5cm]{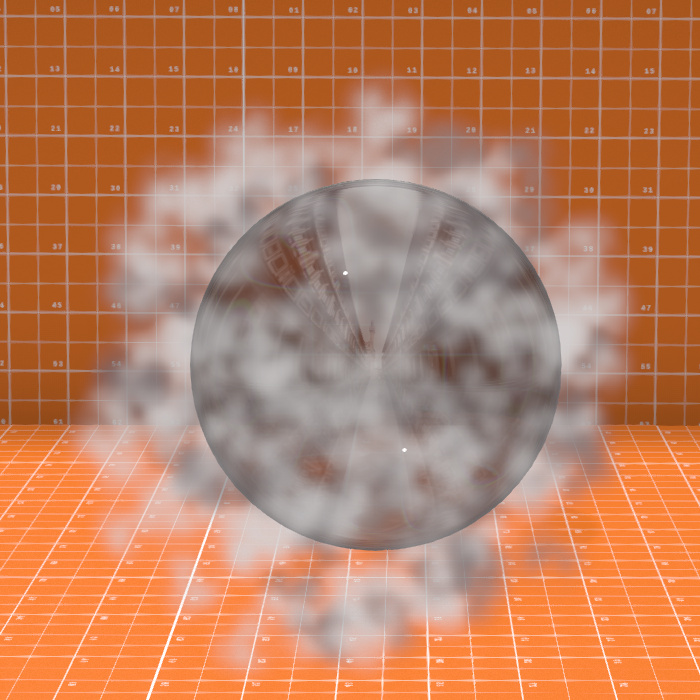}} &
\fbox{\includegraphics[width=0.19\textwidth,clip,trim=0 1cm 0 2.5cm]{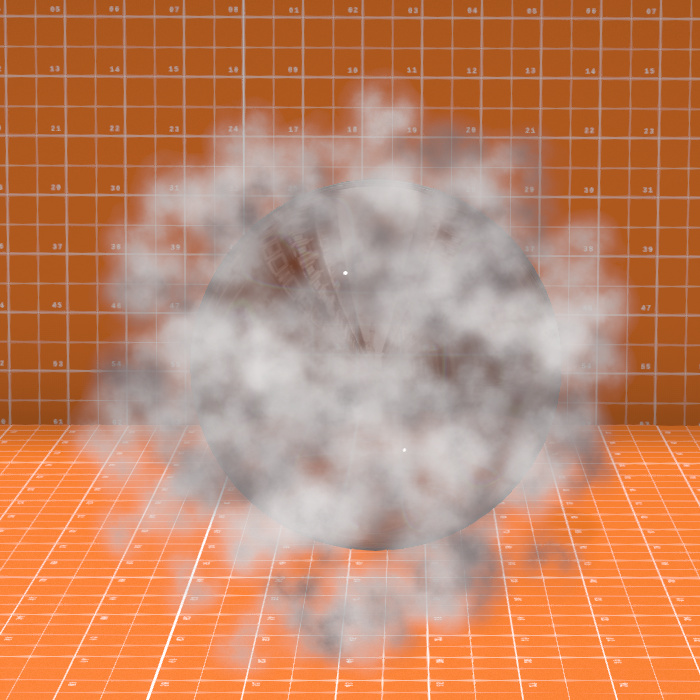}} &
\fbox{\includegraphics[width=0.19\textwidth,clip,trim=0 1cm 0 2.5cm]{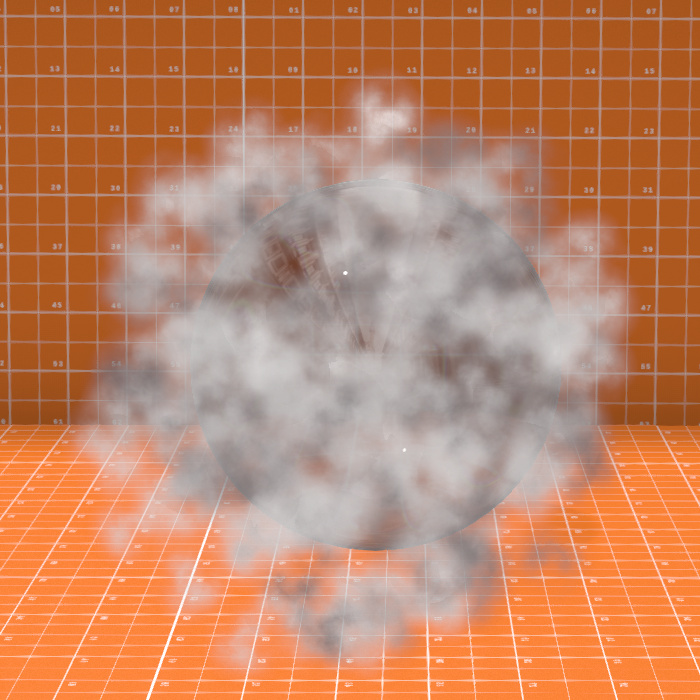}} &
\fbox{\includegraphics[width=0.19\textwidth,clip,trim=0 1cm 0 2.5cm]{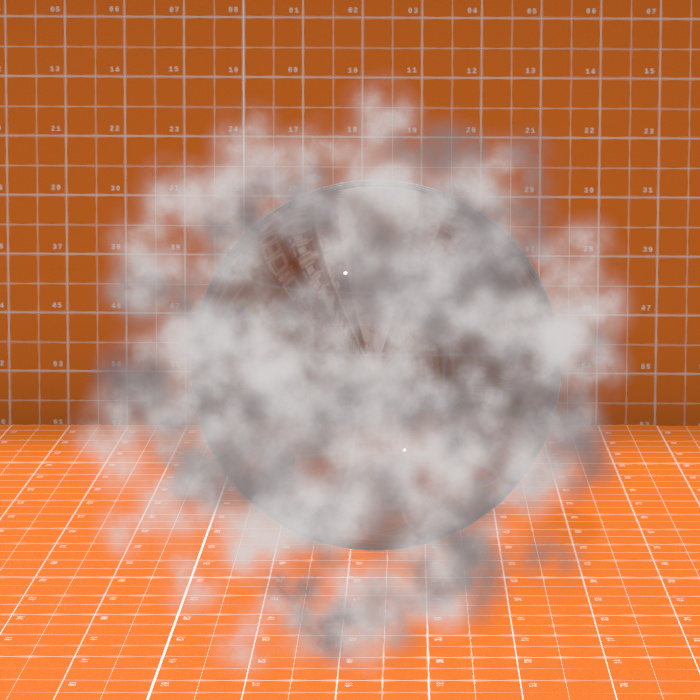}} \\

\!\!\!3\!\!\! &
\fbox{\includegraphics[width=0.19\textwidth]{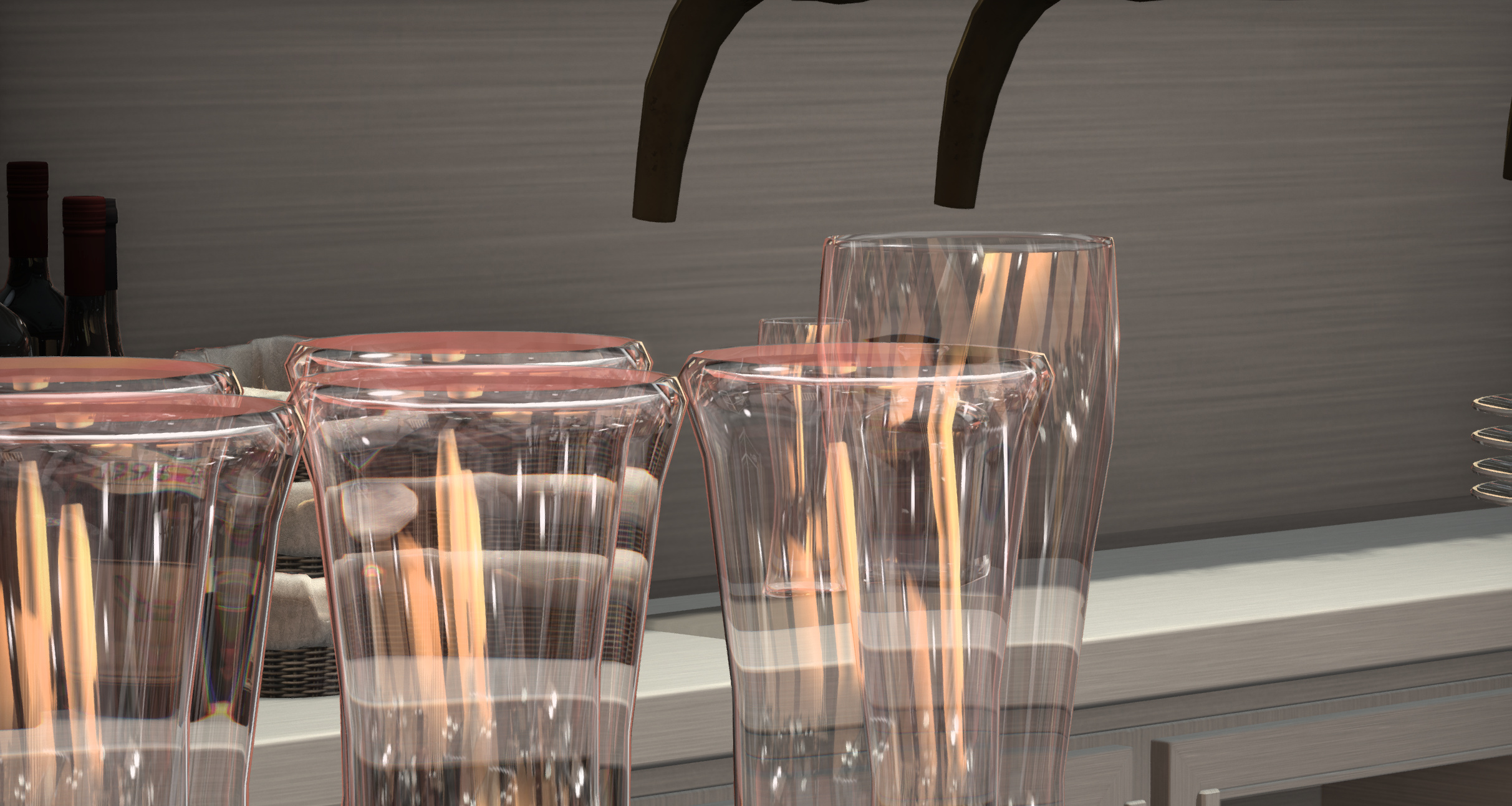}} &
\fbox{\includegraphics[width=0.19\textwidth]{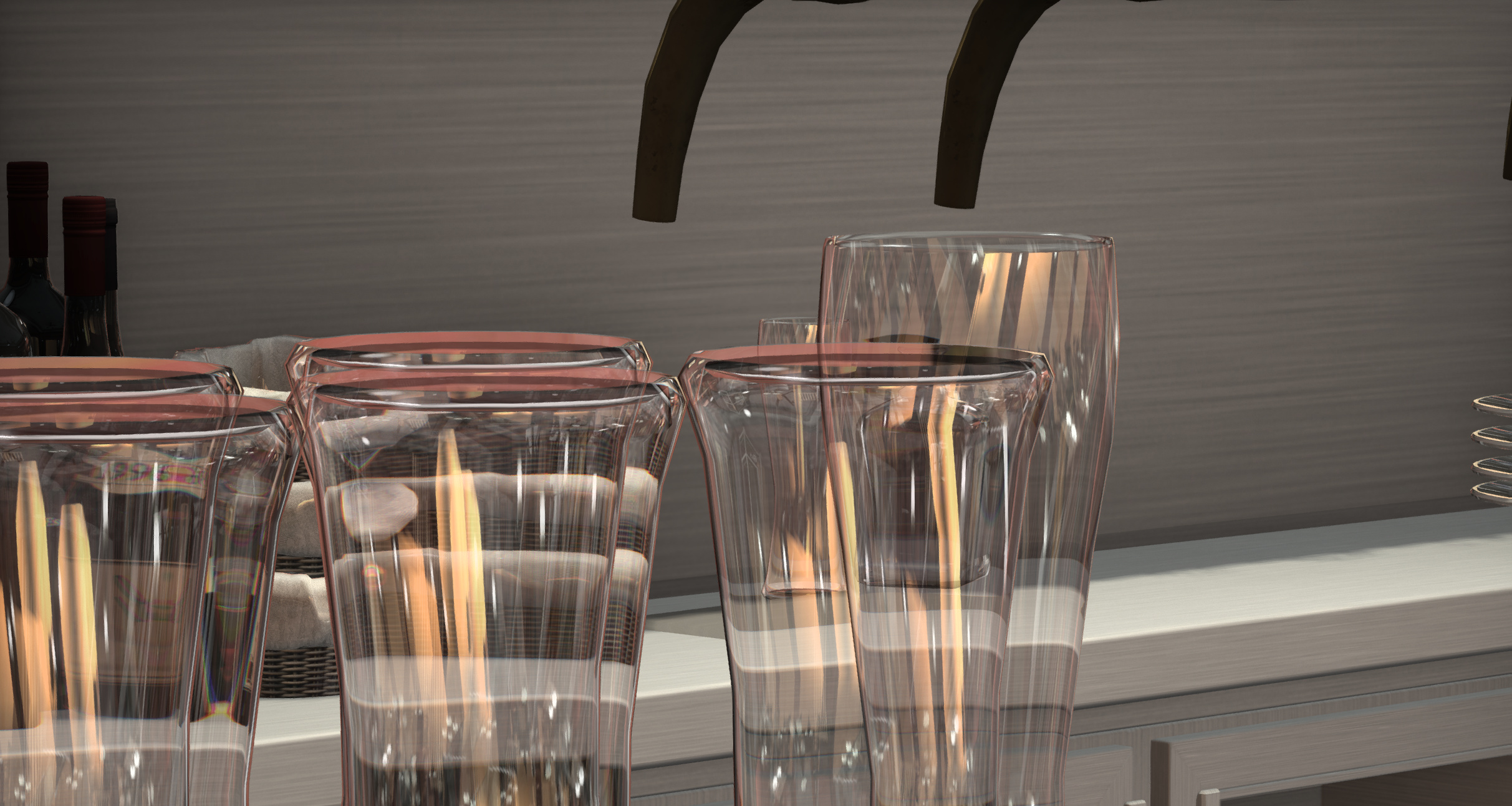}} &
\fbox{\includegraphics[width=0.19\textwidth]{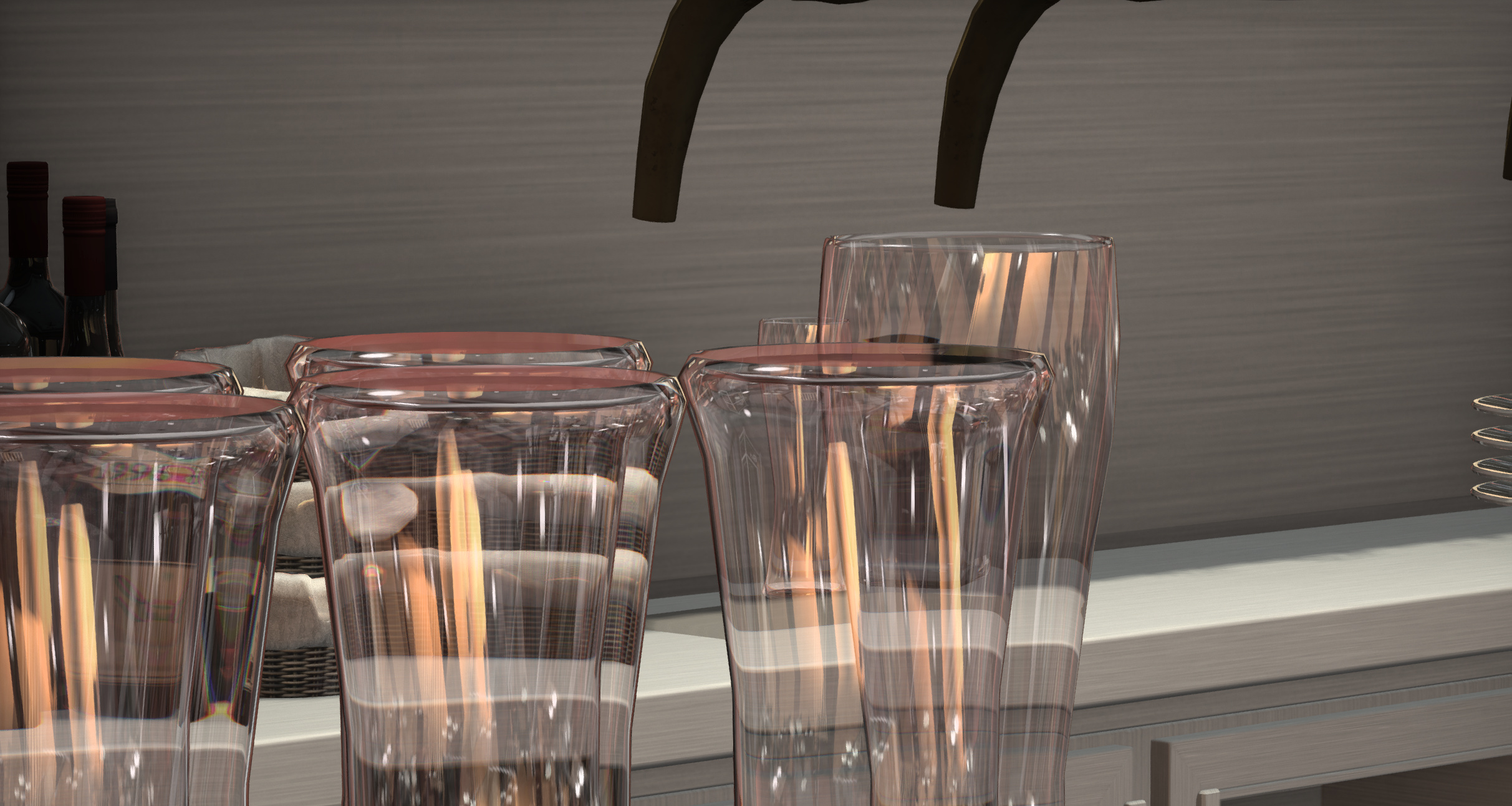}} &
\fbox{\includegraphics[width=0.19\textwidth]{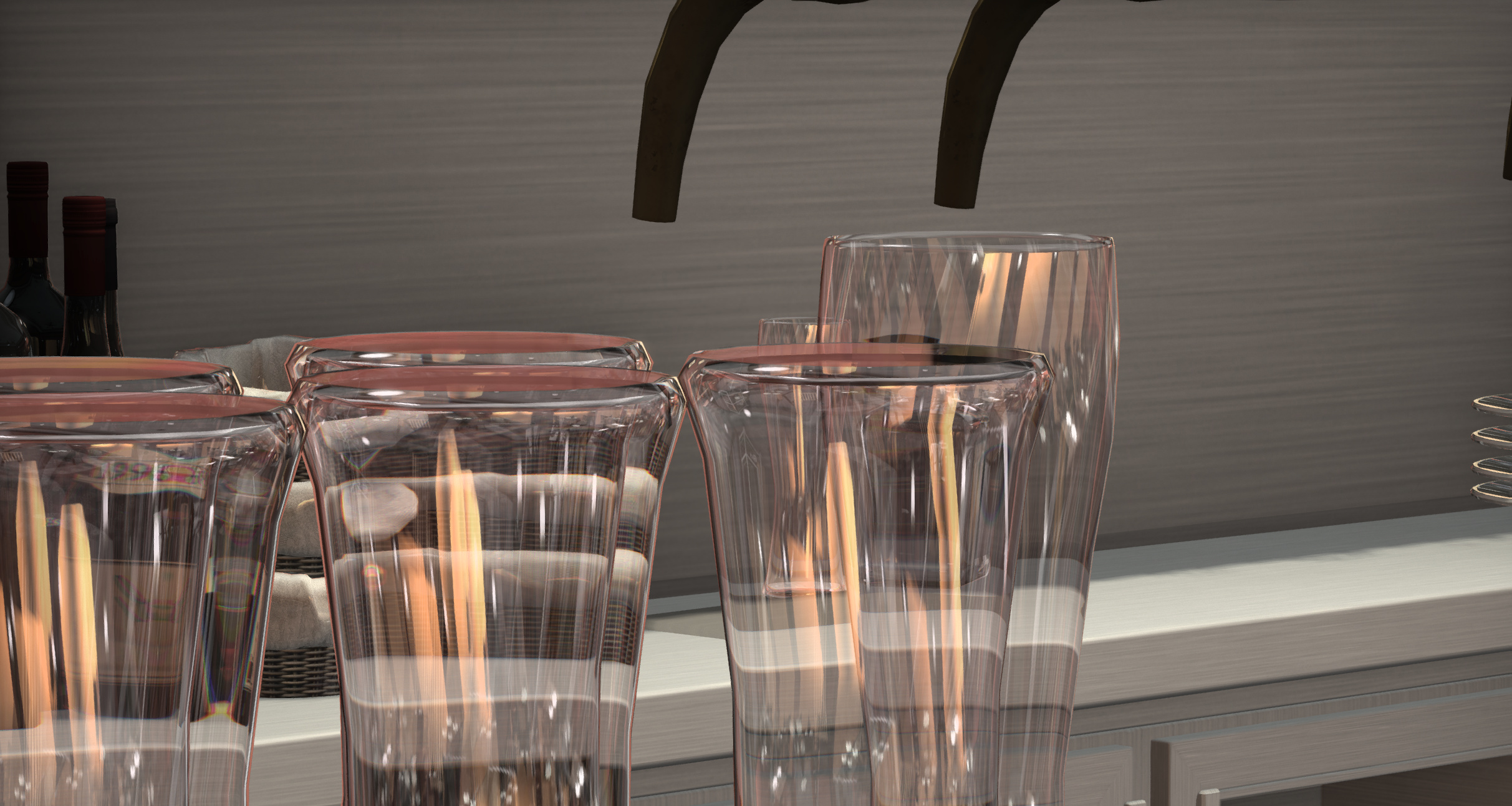}} &
\fbox{\includegraphics[width=0.19\textwidth]{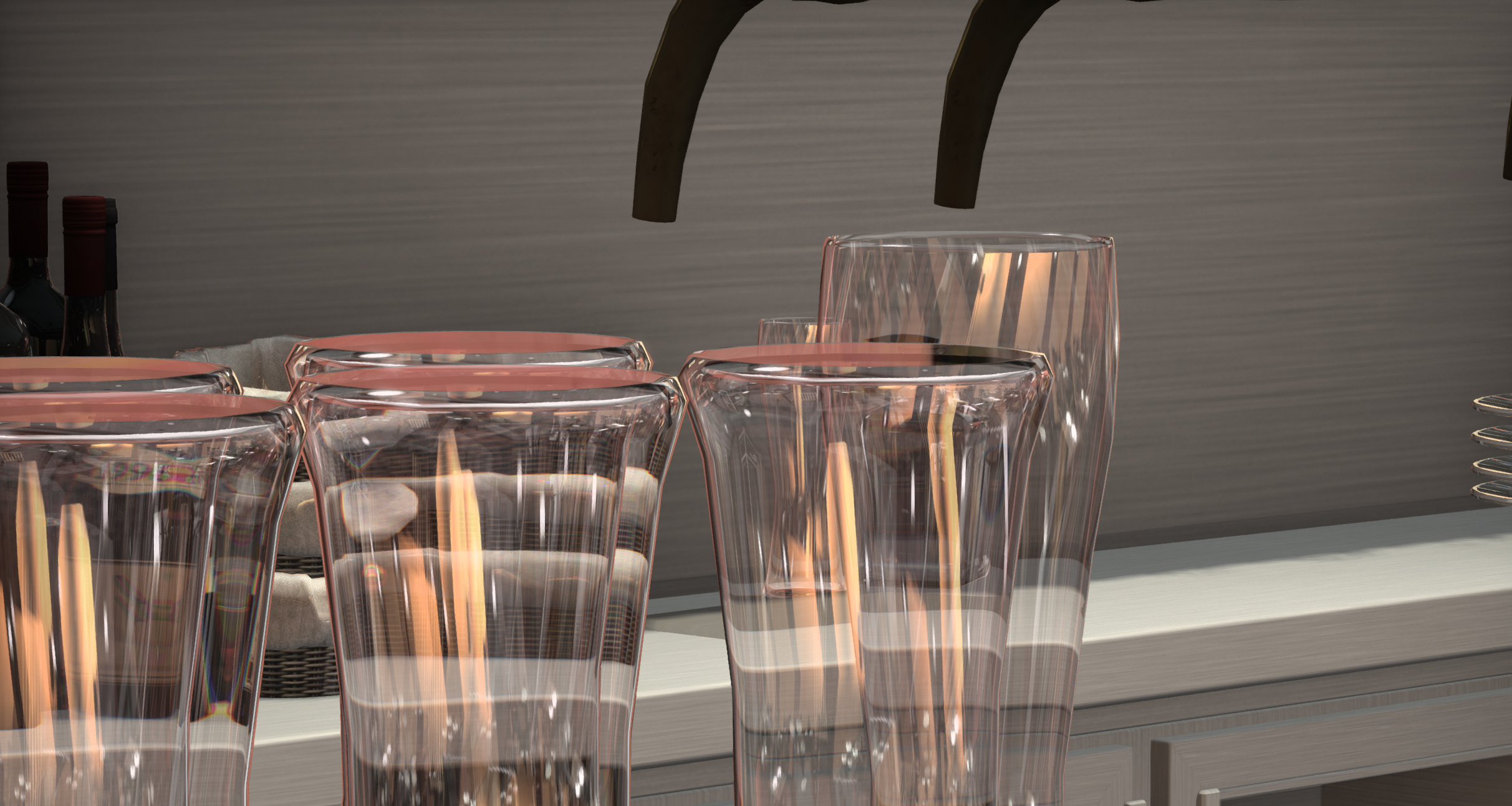}} \\

\!\!\!4\!\!\! &
\fbox{\includegraphics[width=0.19\textwidth]{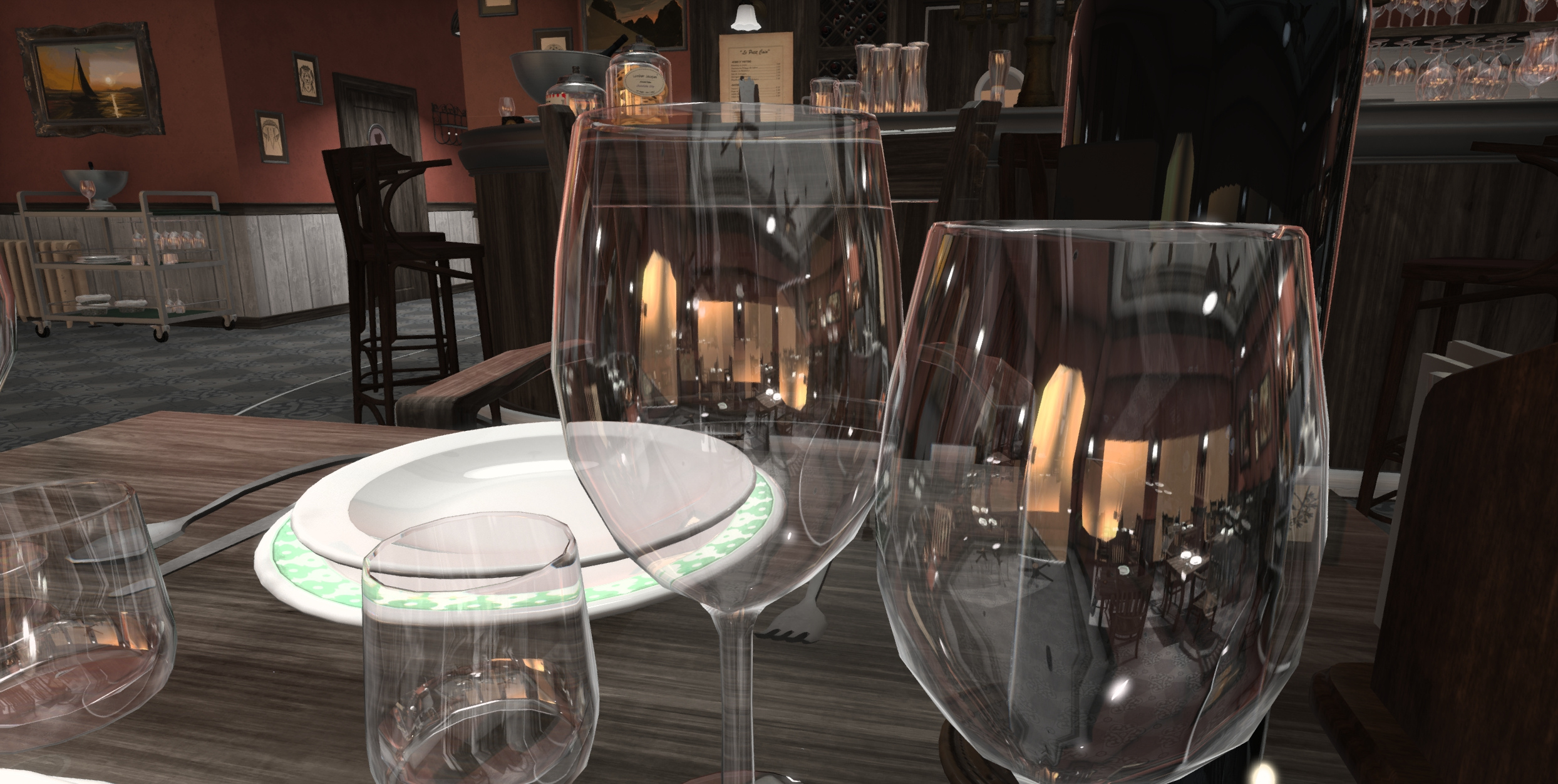}} &
\fbox{\includegraphics[width=0.19\textwidth]{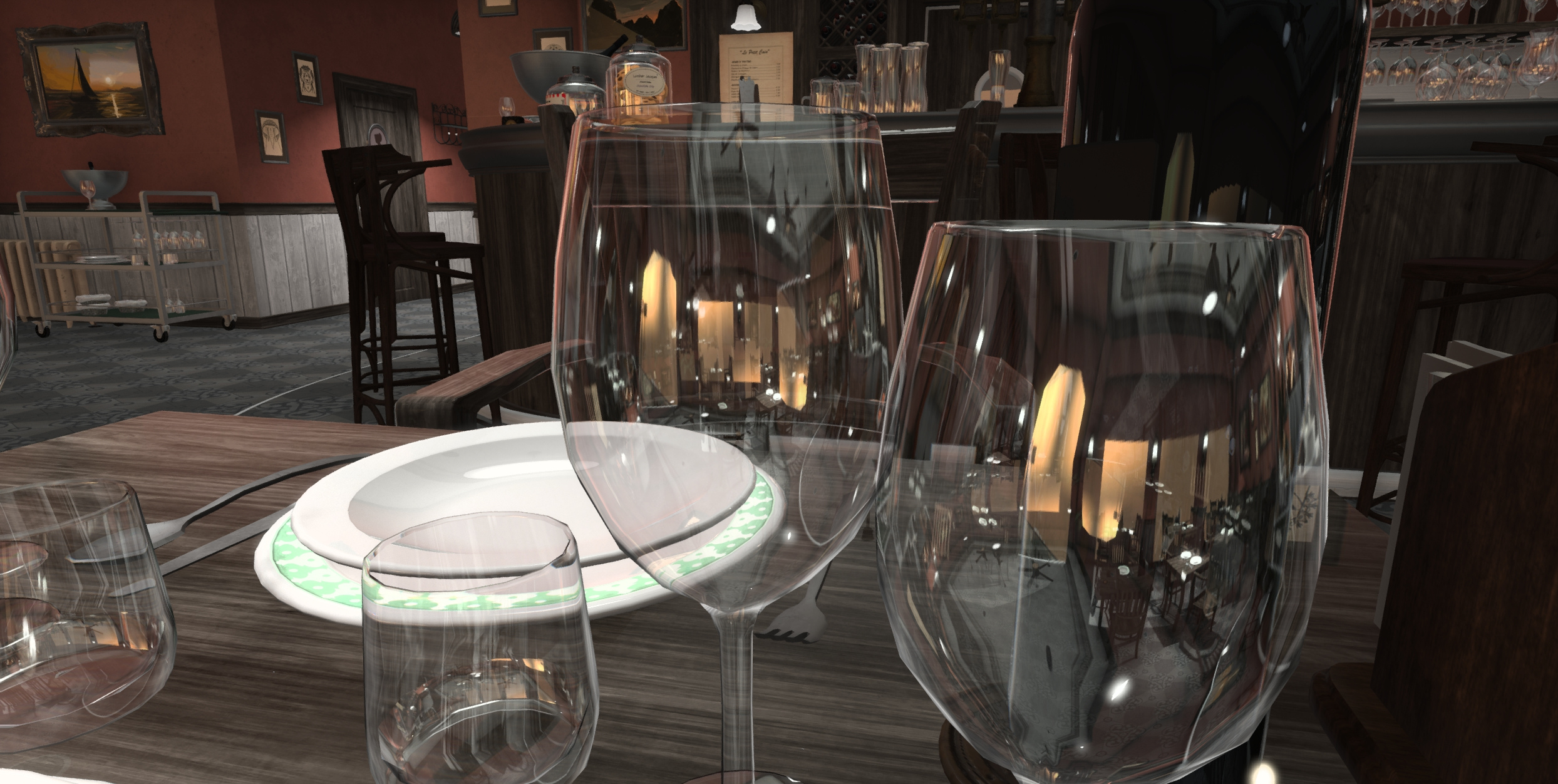}} &
\fbox{\includegraphics[width=0.19\textwidth]{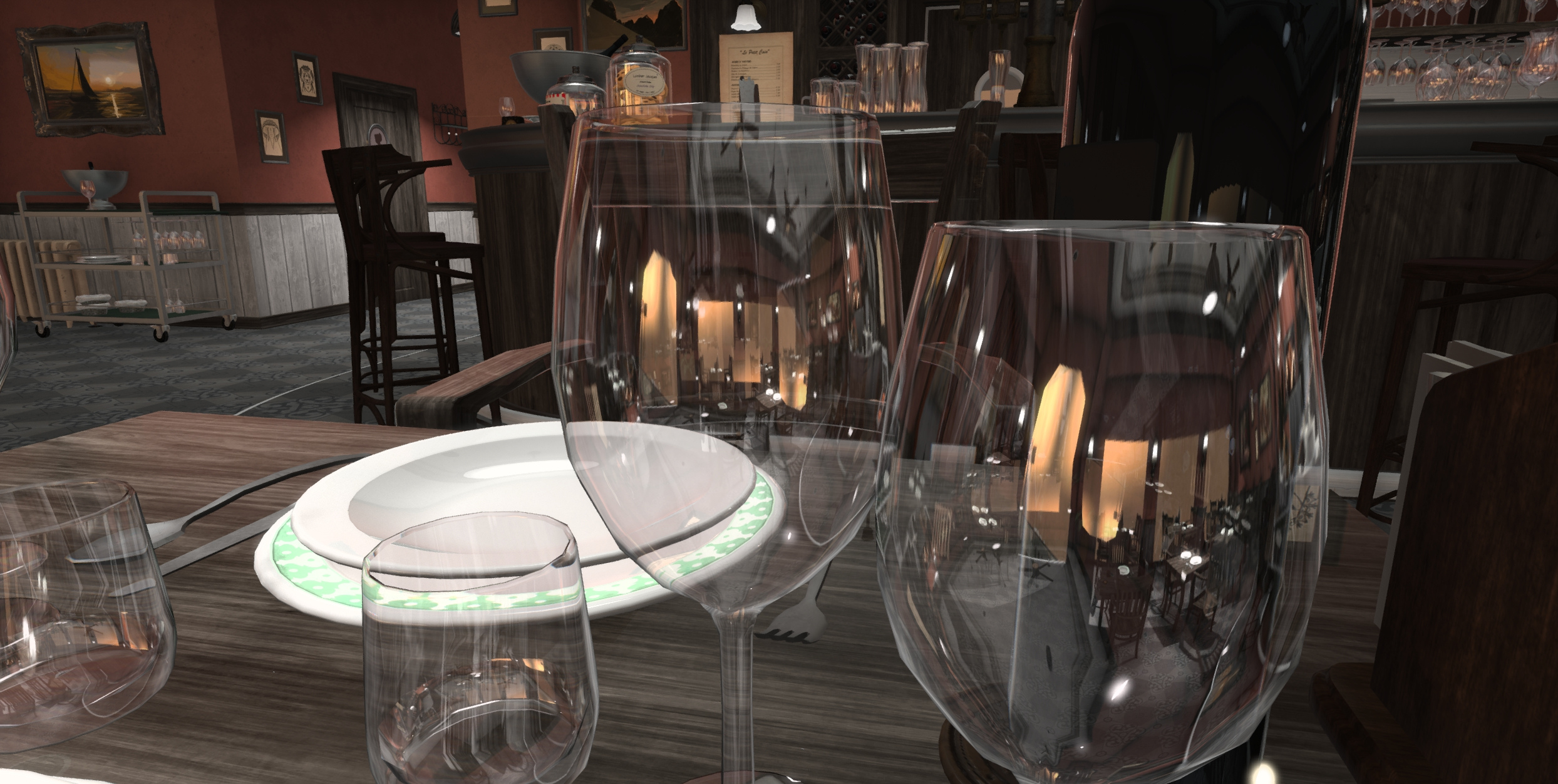}} &
\fbox{\includegraphics[width=0.19\textwidth]{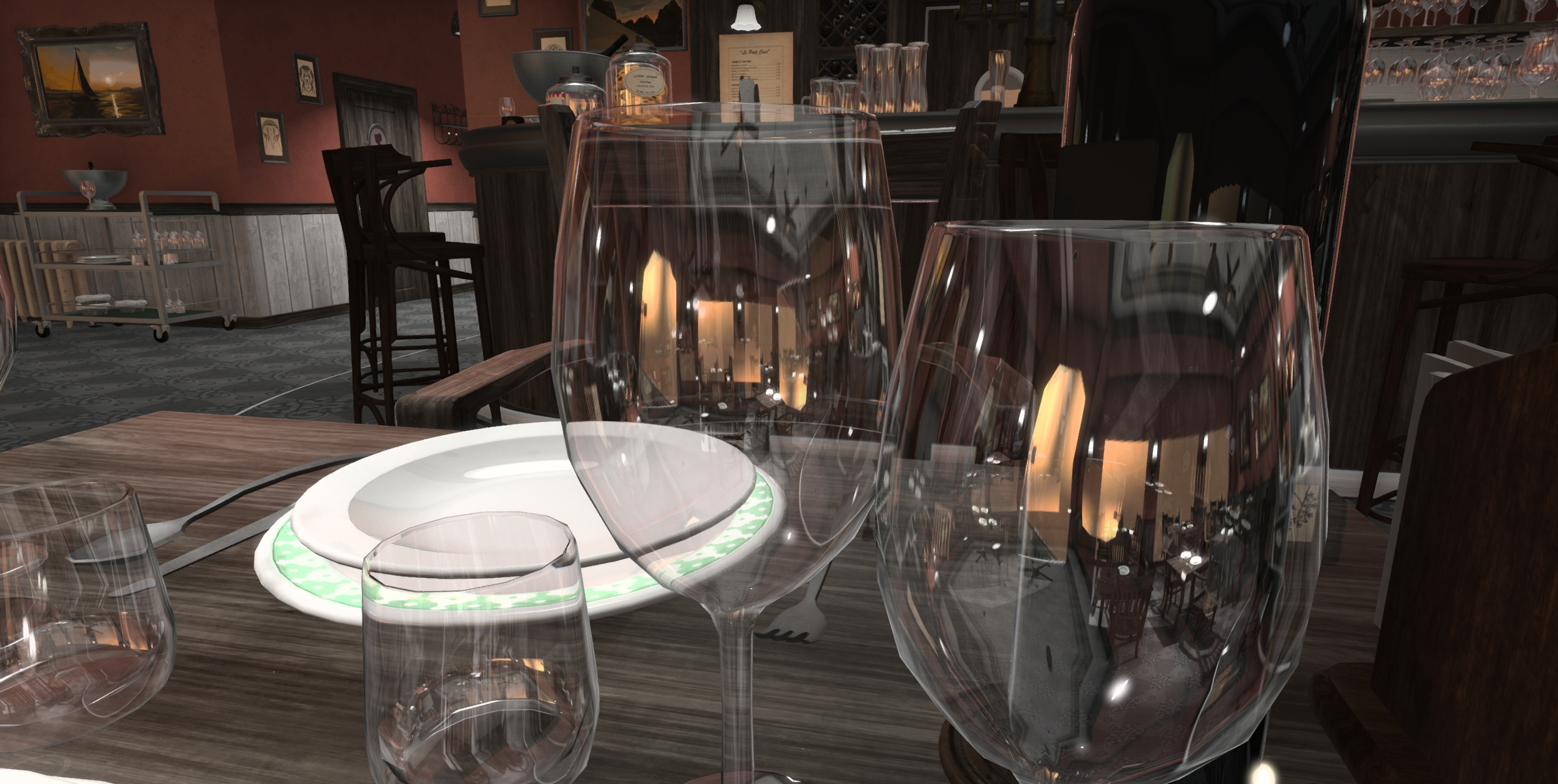}} &
\fbox{\includegraphics[width=0.19\textwidth]{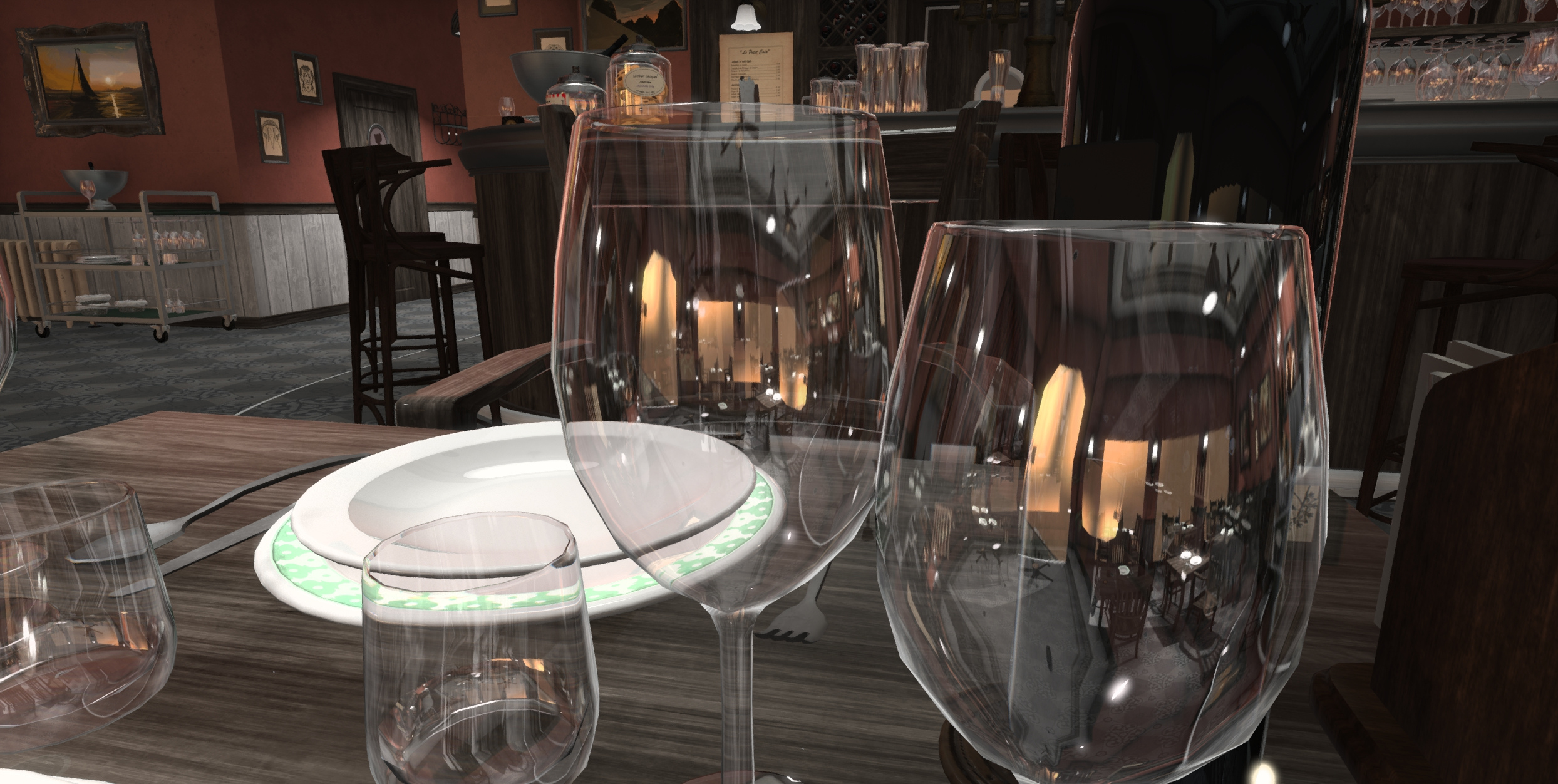}} \\

\!\!\!5\!\!\! &
\fbox{\includegraphics[width=0.19\textwidth]{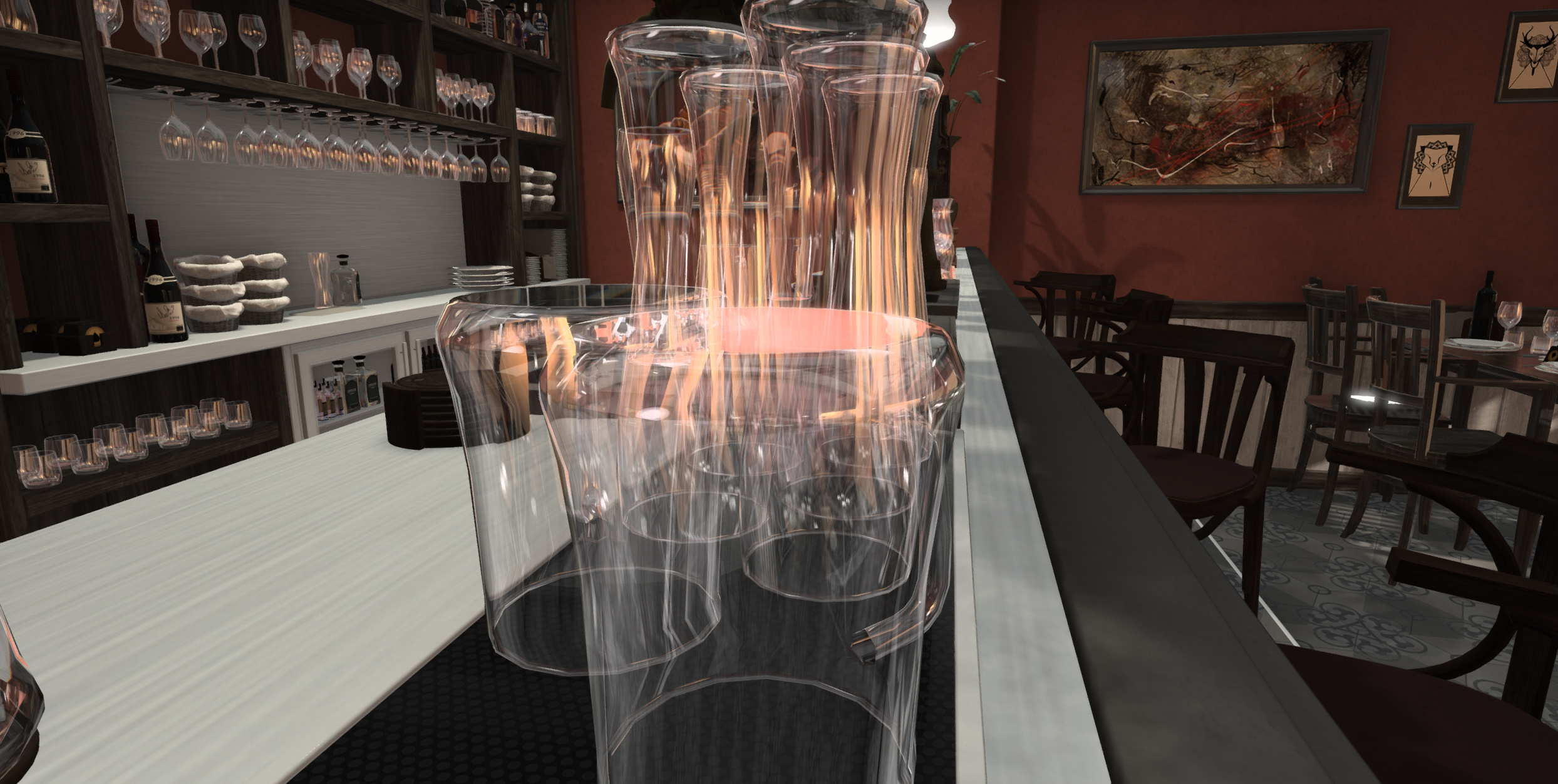}} &
\fbox{\includegraphics[width=0.19\textwidth]{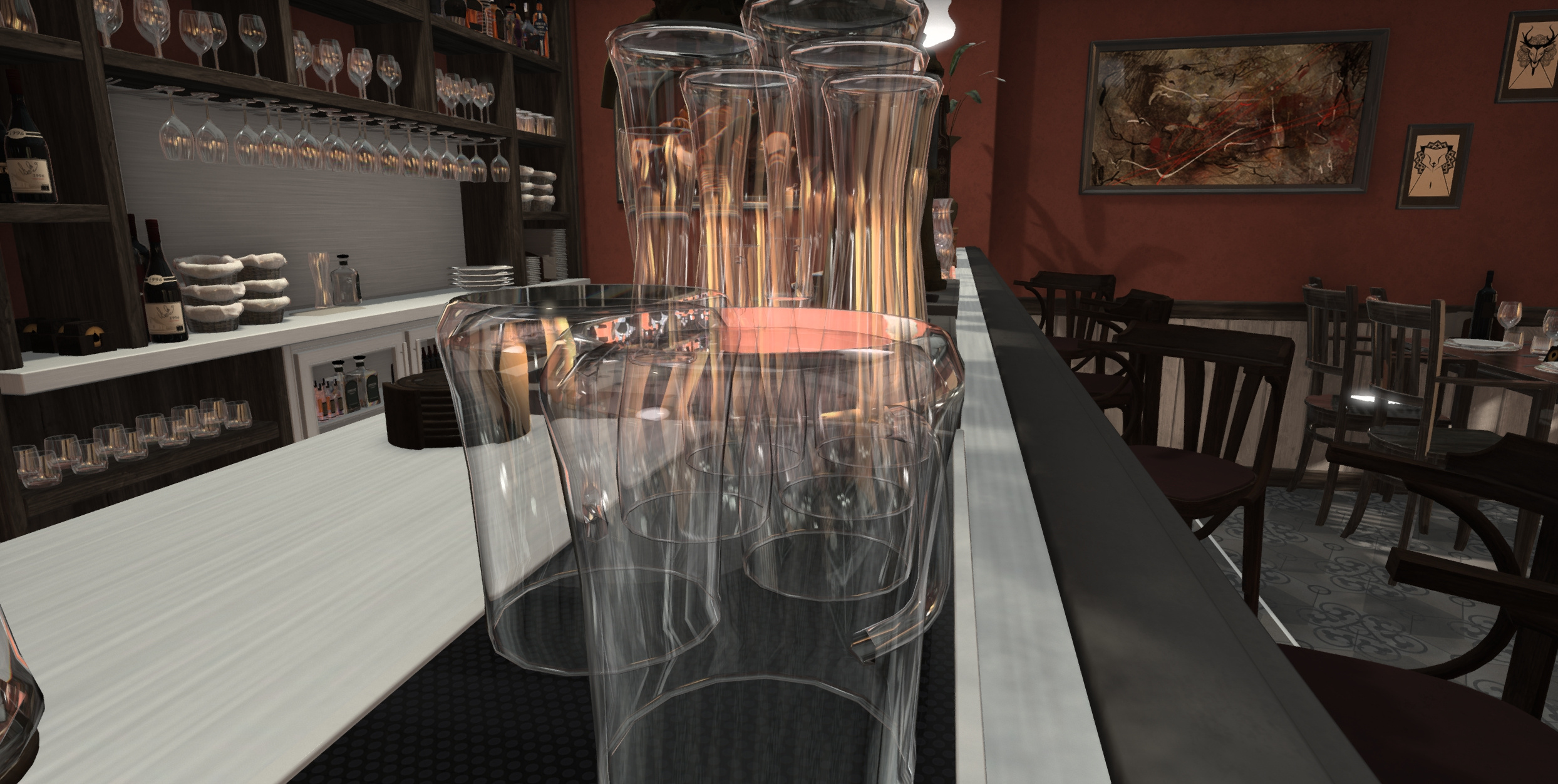}} &
\fbox{\includegraphics[width=0.19\textwidth]{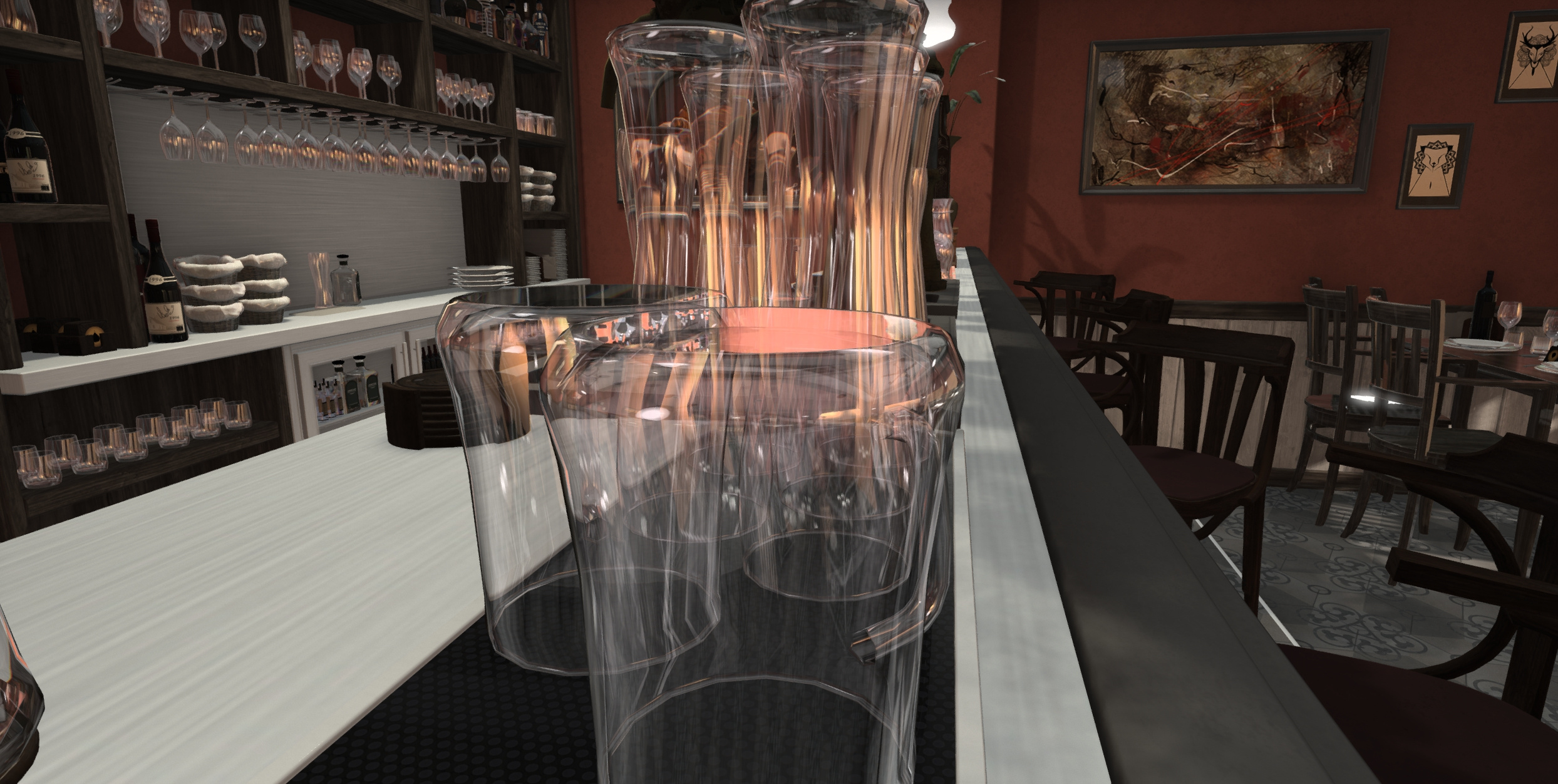}} &
\fbox{\includegraphics[width=0.19\textwidth]{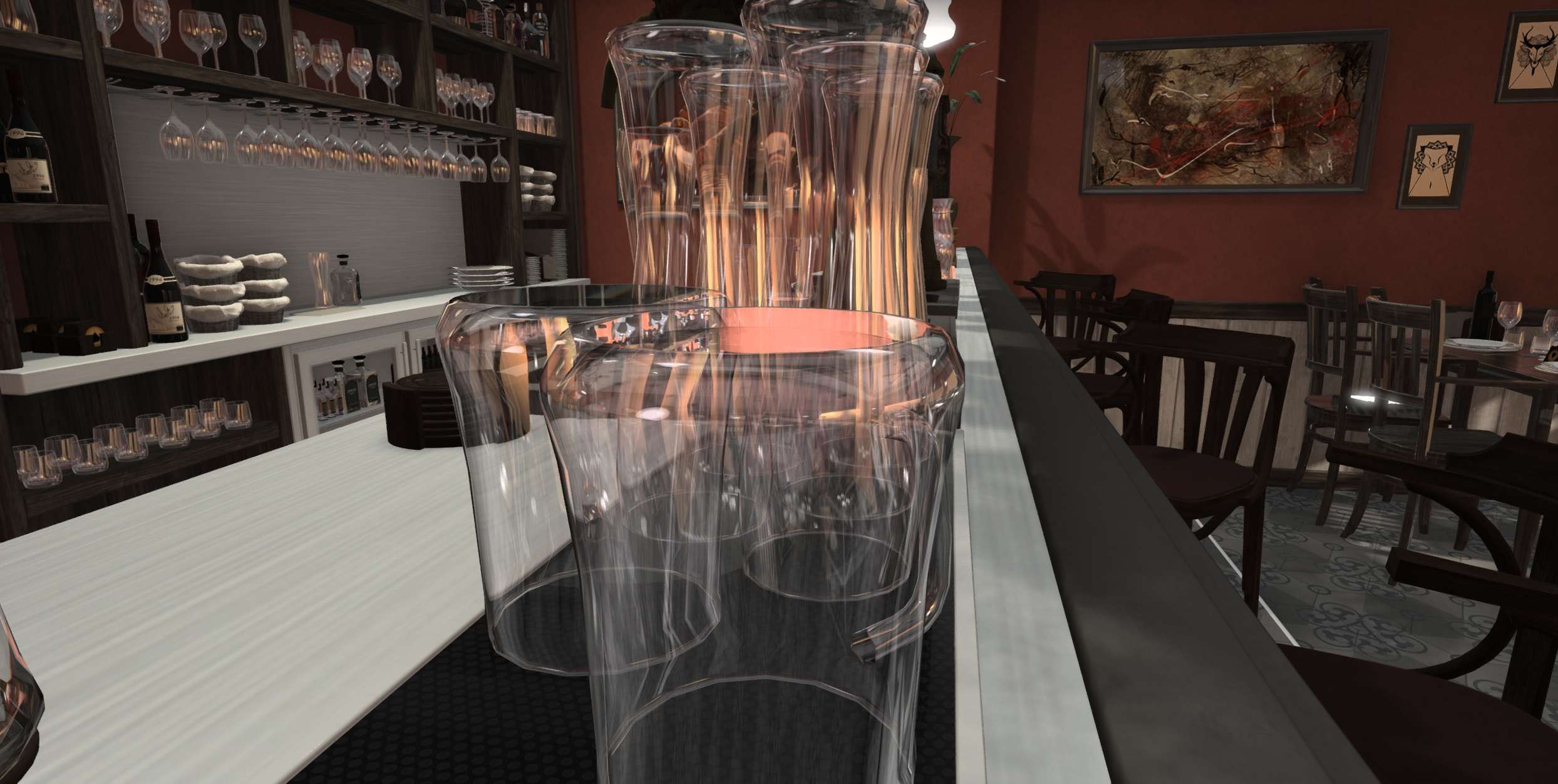}} &
\fbox{\includegraphics[width=0.19\textwidth]{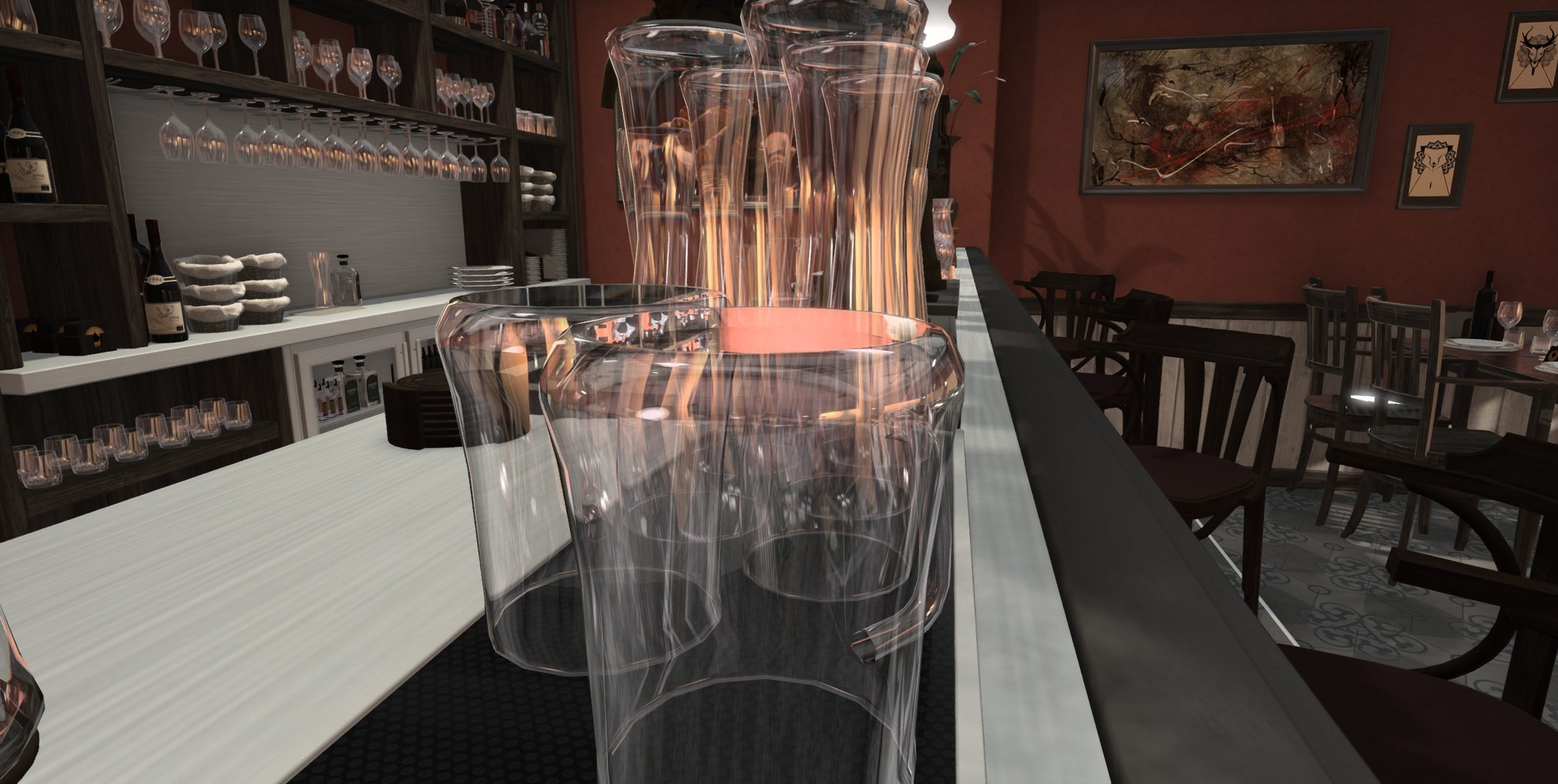}}\\

\!\!\!6\!\!\! &
\fbox{\includegraphics[width=0.19\textwidth]{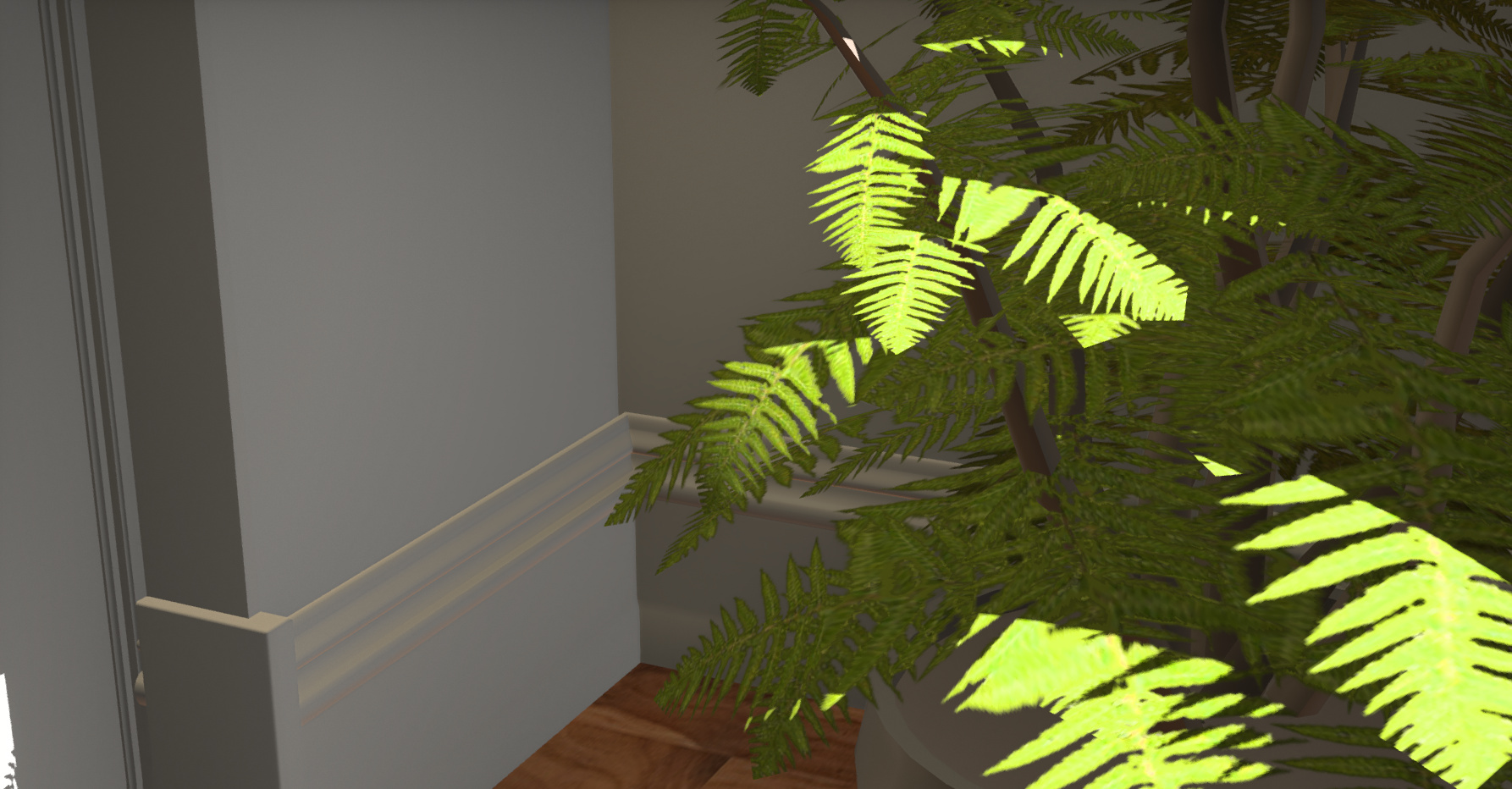}} &
\fbox{\includegraphics[width=0.19\textwidth]{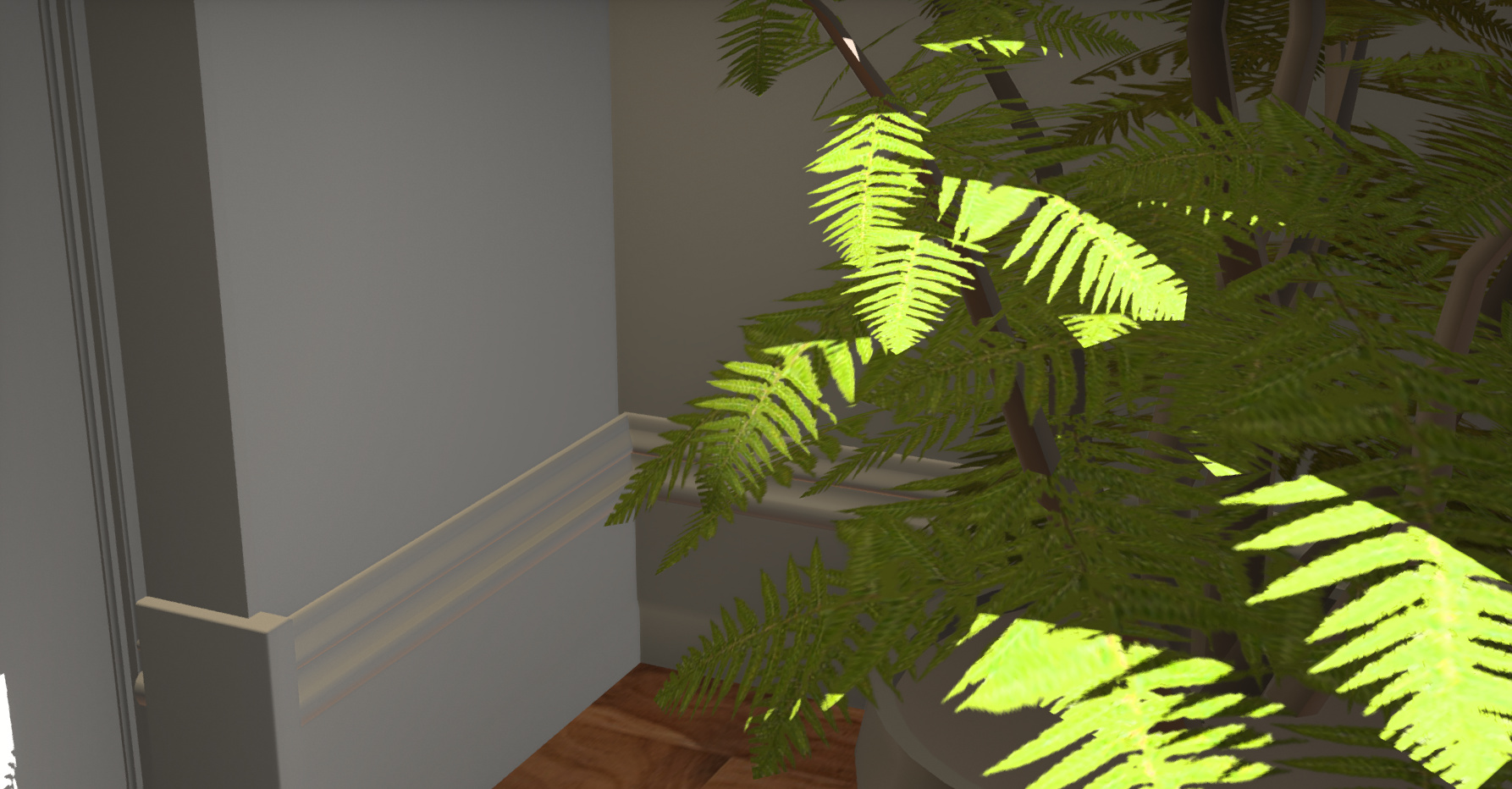}} &
\fbox{\includegraphics[width=0.19\textwidth]{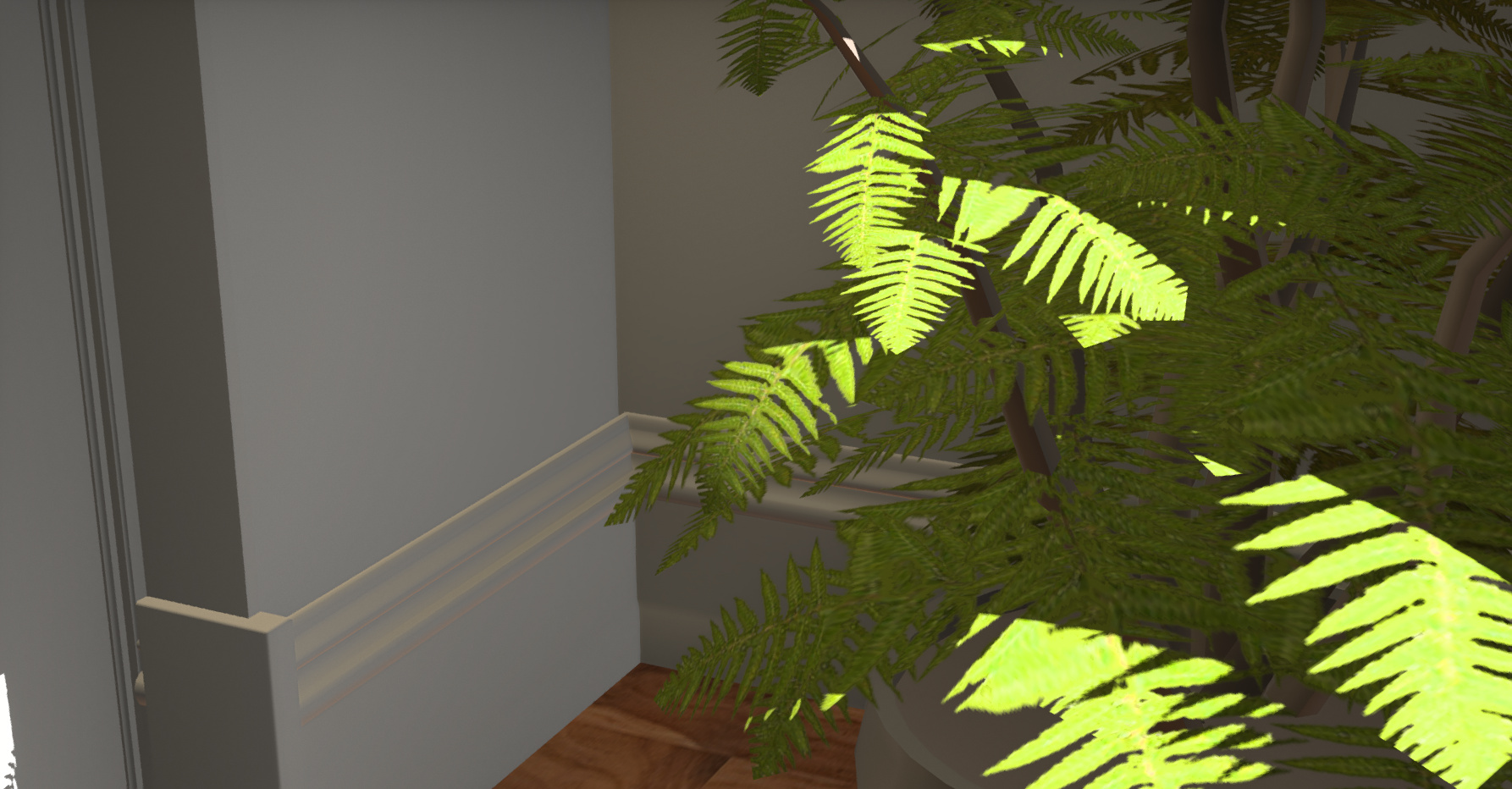}} &
\fbox{\includegraphics[width=0.19\textwidth]{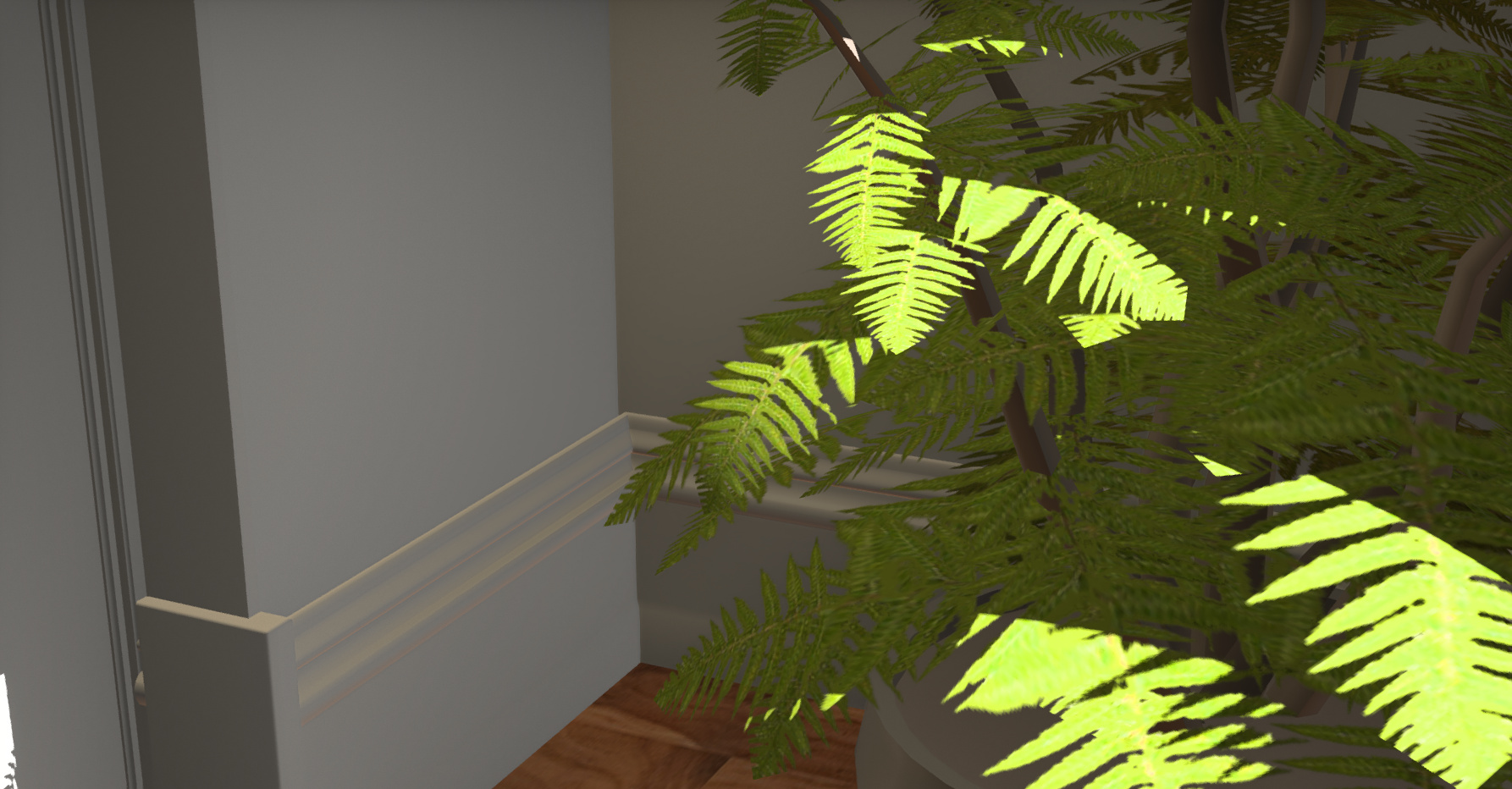}} &
\fbox{\includegraphics[width=0.19\textwidth]{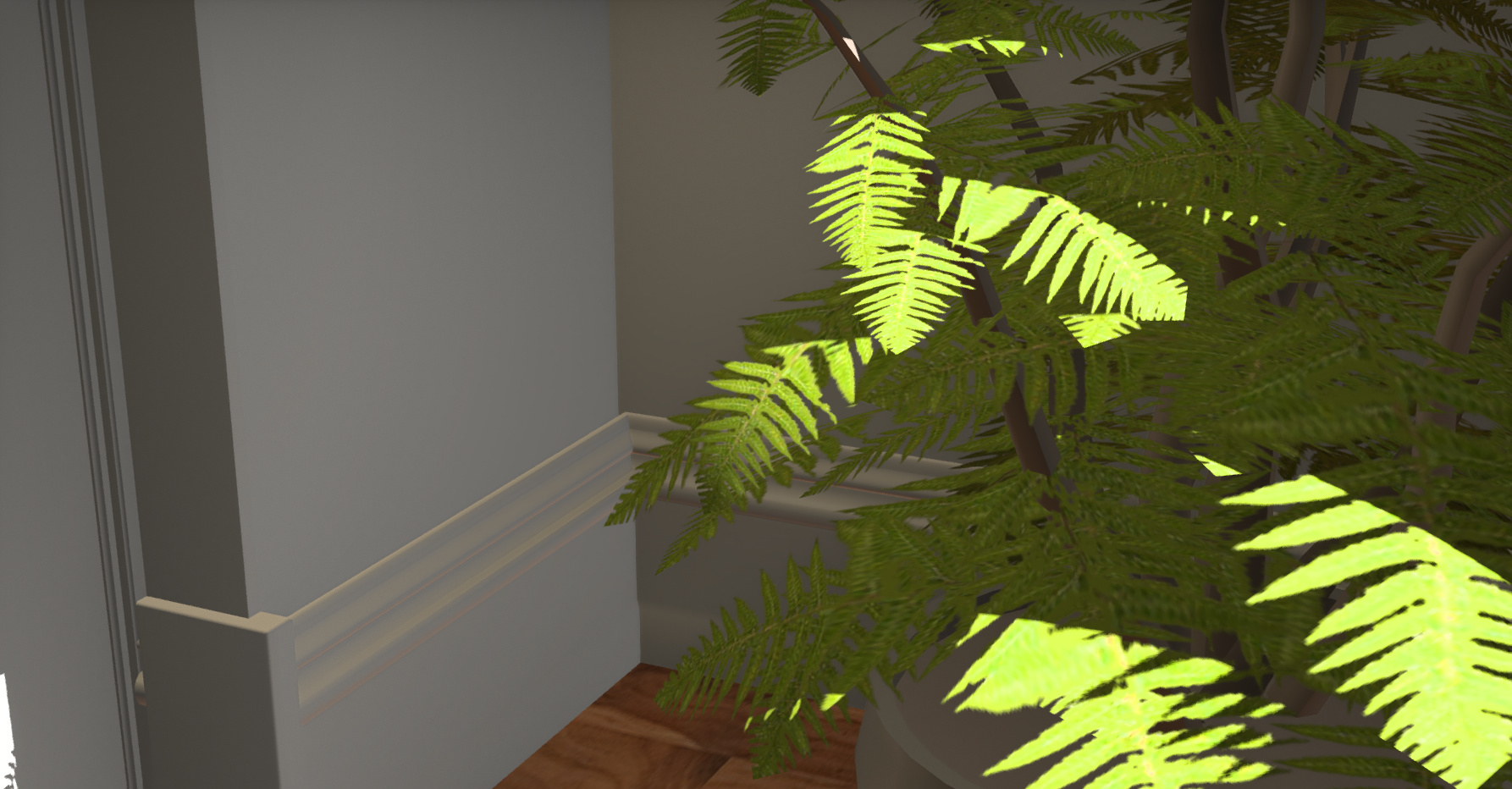}} \\

\!\!\!7\!\!\! &
\fbox{\includegraphics[width=0.19\textwidth]{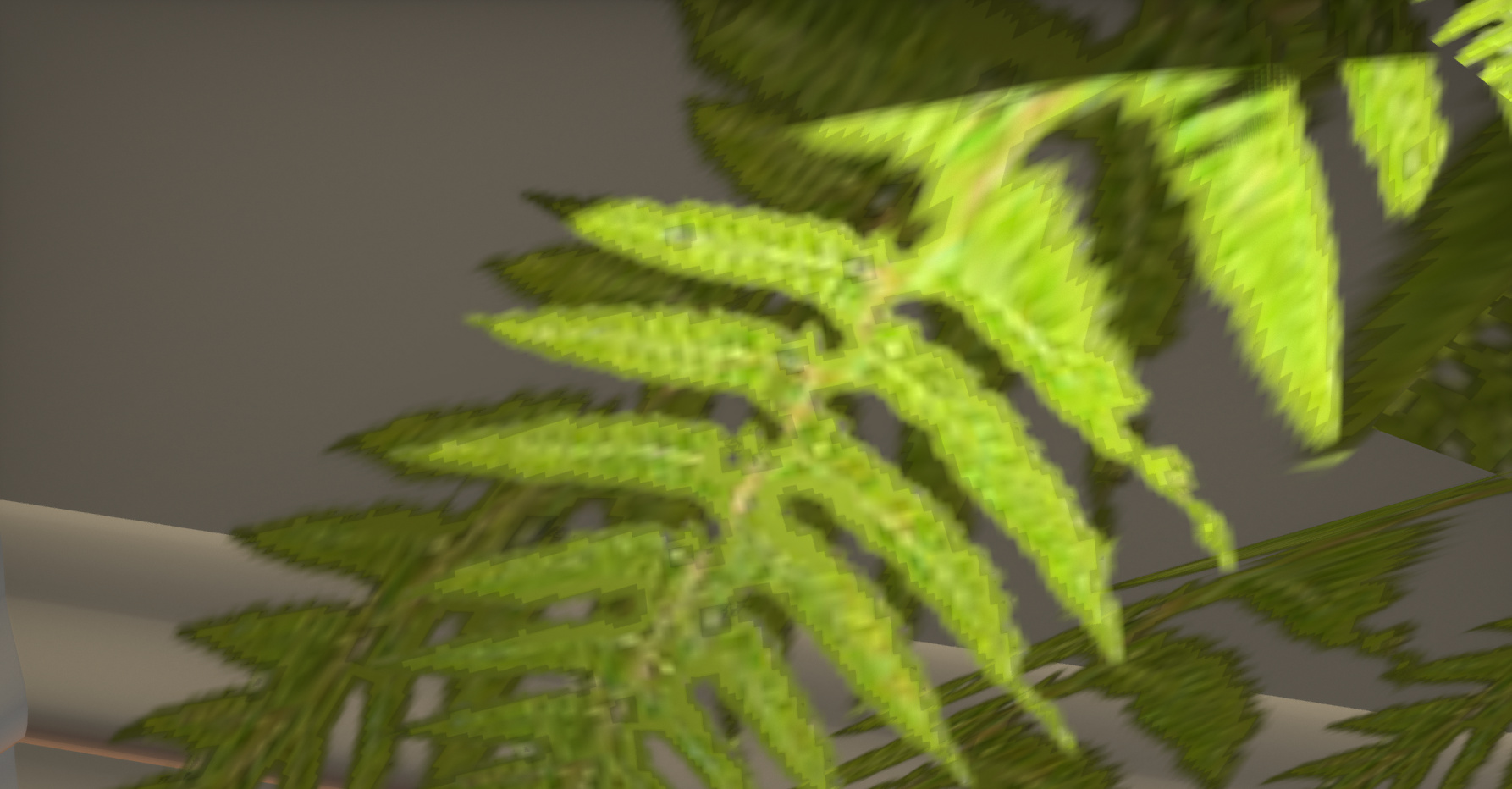}} &
\fbox{\includegraphics[width=0.19\textwidth]{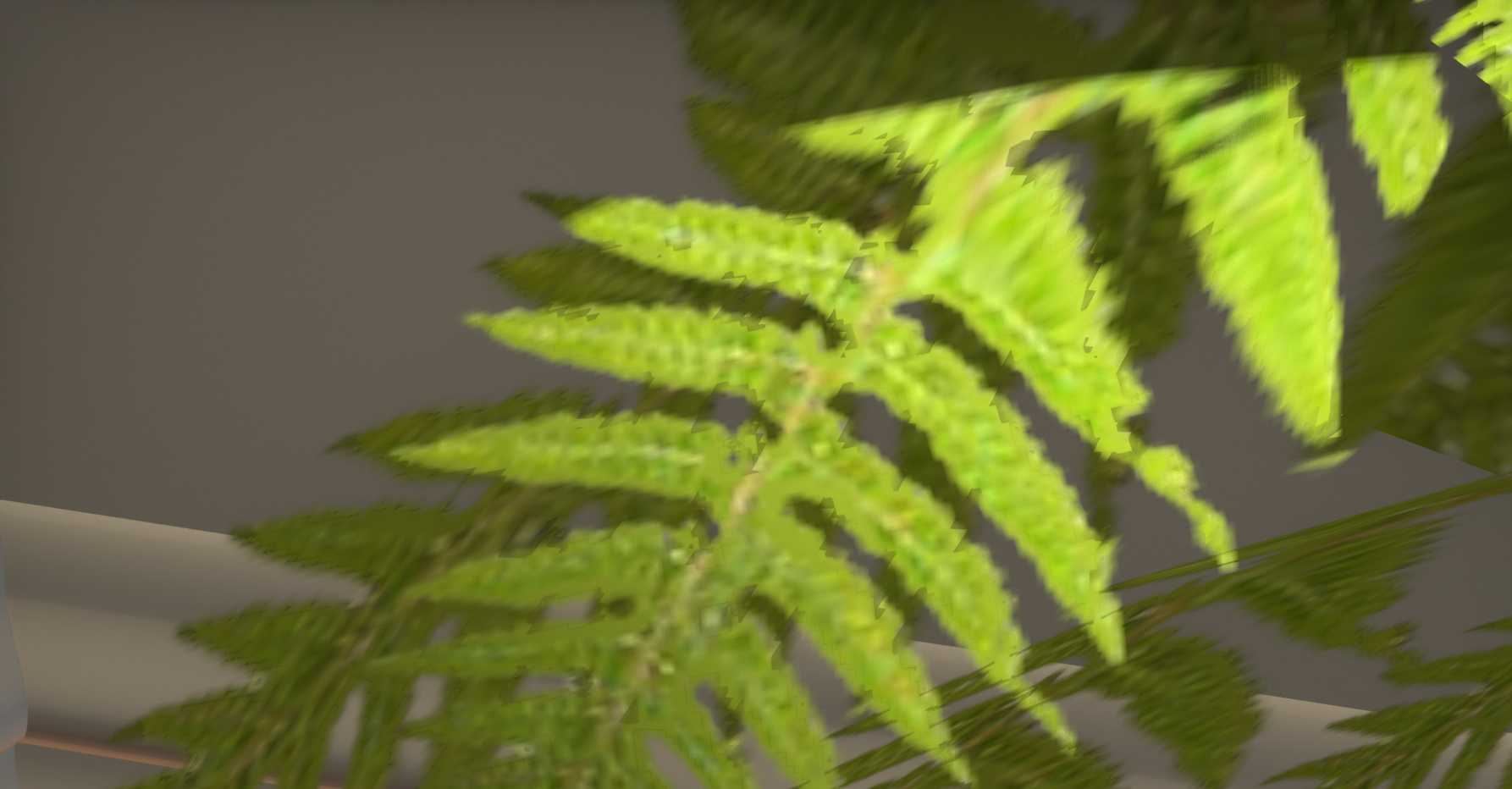}} &
\fbox{\includegraphics[width=0.19\textwidth]{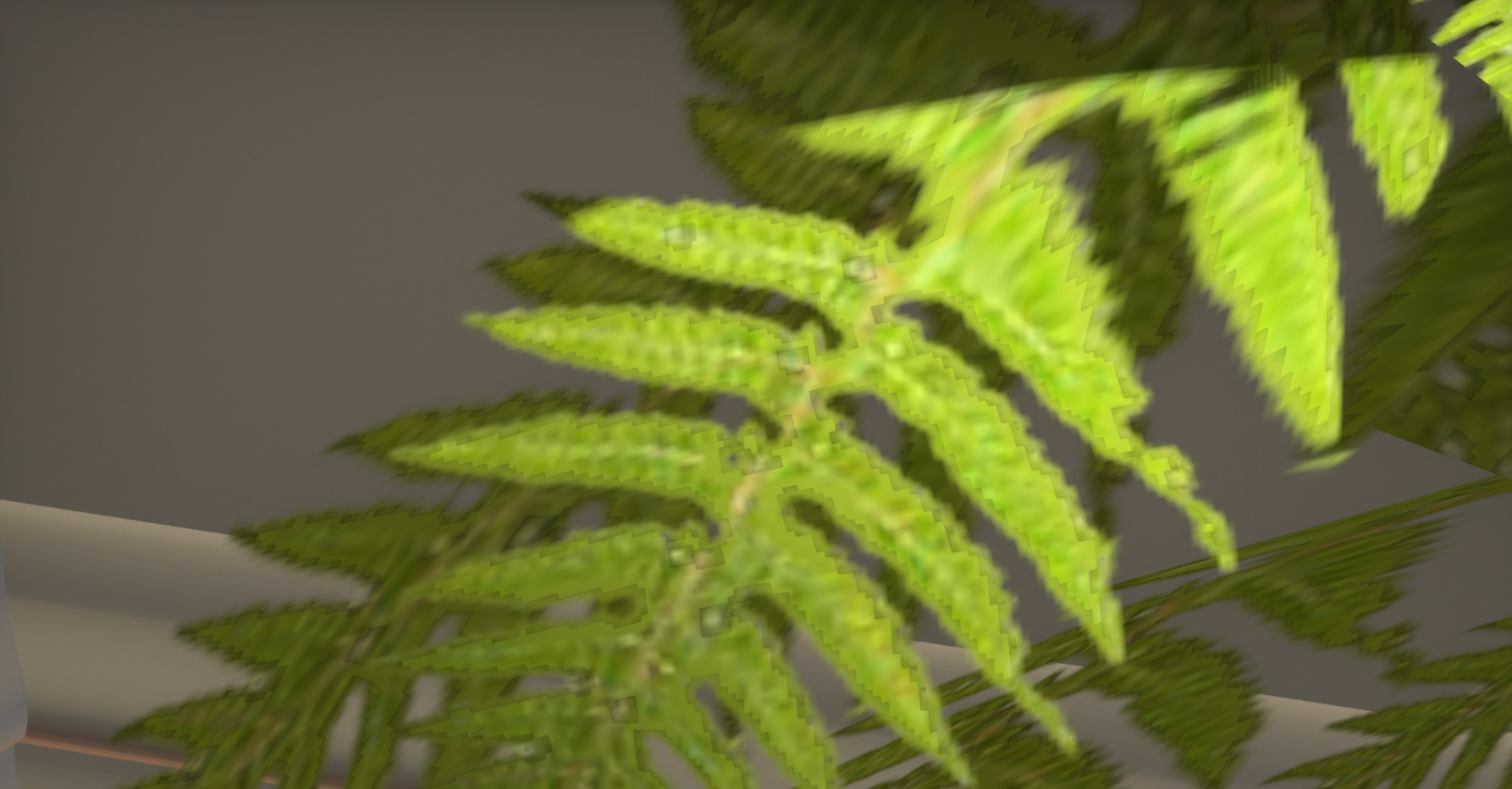}} &
\fbox{\includegraphics[width=0.19\textwidth]{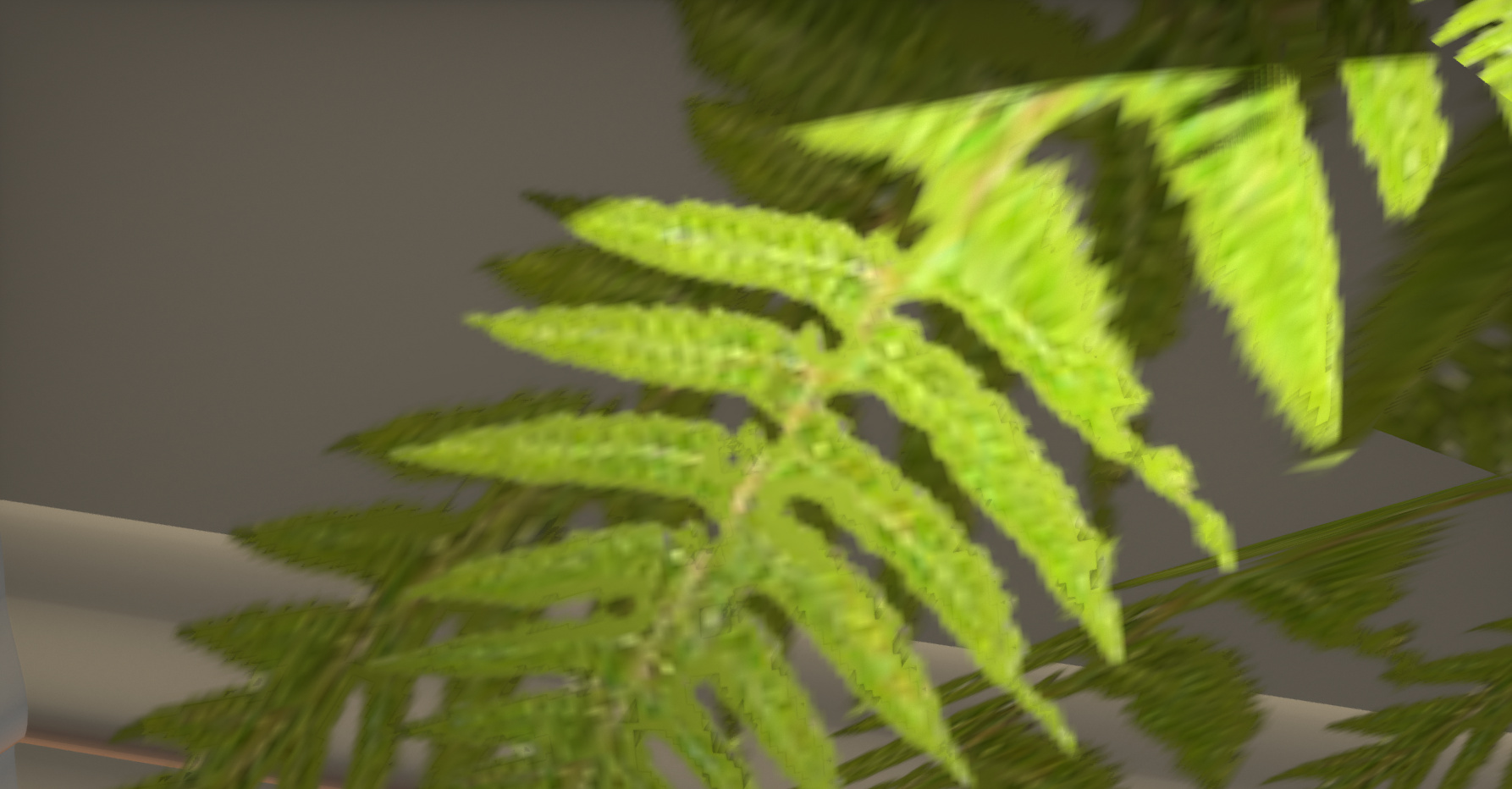}} &
\fbox{\includegraphics[width=0.19\textwidth]{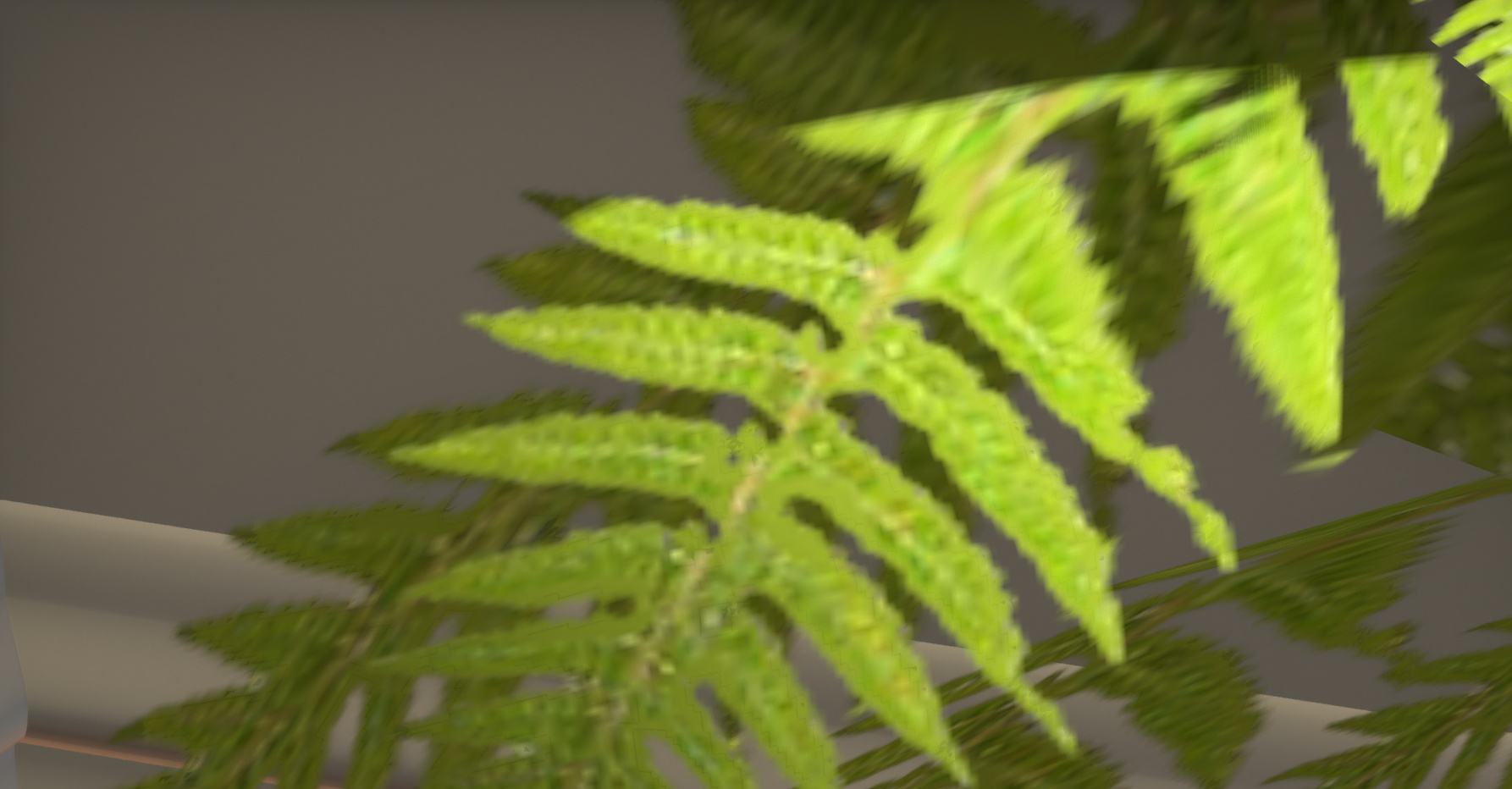}} \\

\!\!\!8\!\!\! &
\fbox{\includegraphics[width=0.19\textwidth]{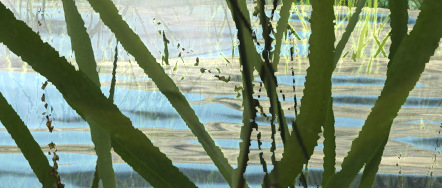}} &
\fbox{\includegraphics[width=0.19\textwidth]{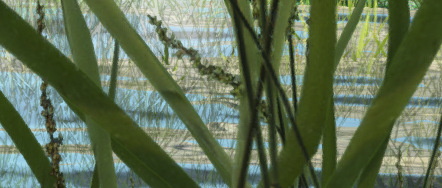}} &
\fbox{\includegraphics[width=0.19\textwidth]{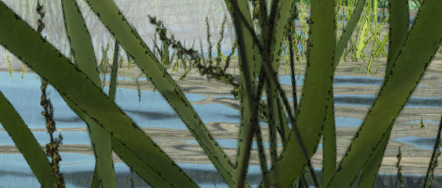}} &
\fbox{\includegraphics[width=0.19\textwidth]{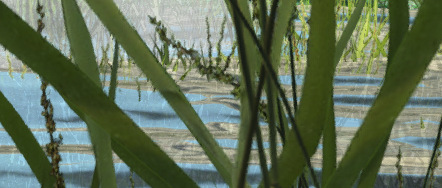}} &
\fbox{\includegraphics[width=0.19\textwidth]{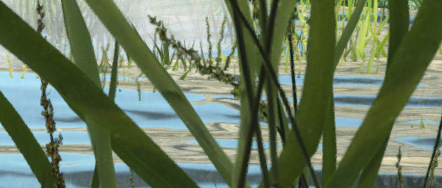}} \\

\!\!\!9\!\!\! &
\fbox{\includegraphics[width=0.19\textwidth]{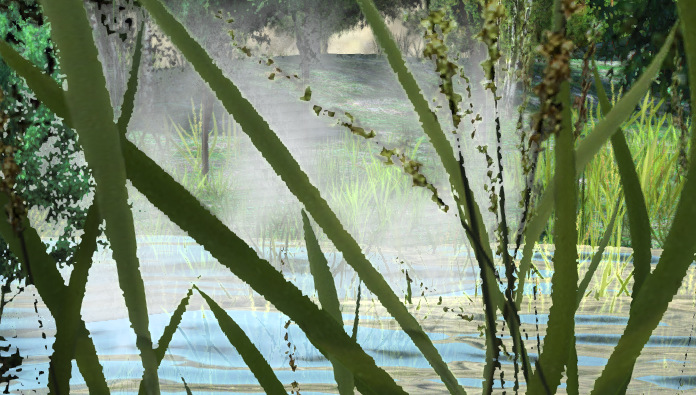}} &
\fbox{\includegraphics[width=0.19\textwidth]{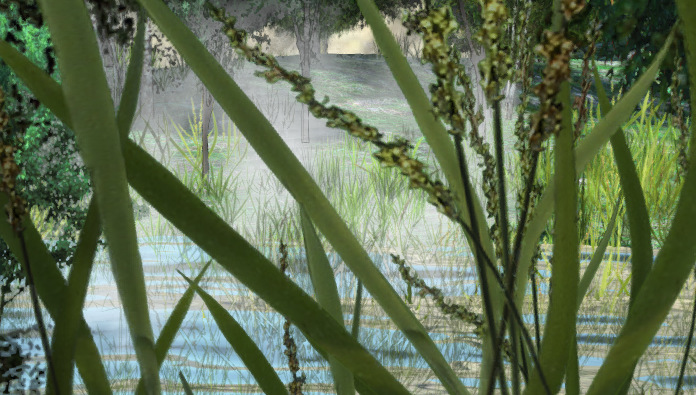}} &
\fbox{\includegraphics[width=0.19\textwidth]{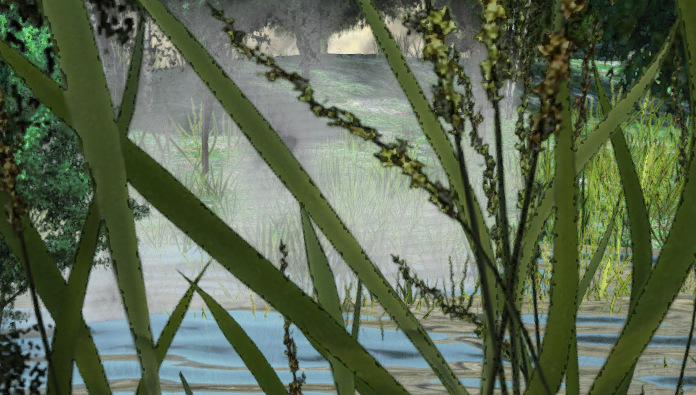}} &
\fbox{\includegraphics[width=0.19\textwidth]{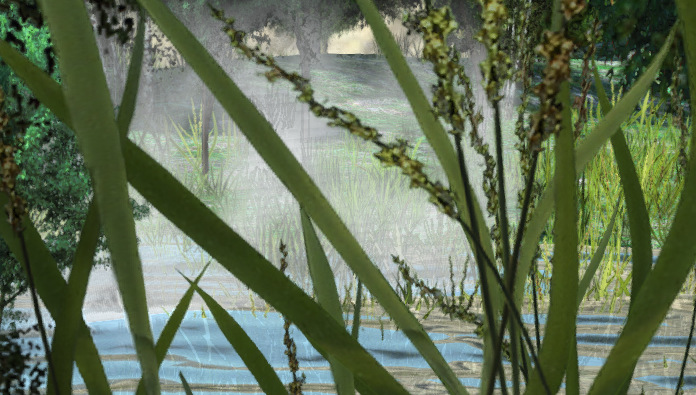}} &
\fbox{\includegraphics[width=0.19\textwidth]{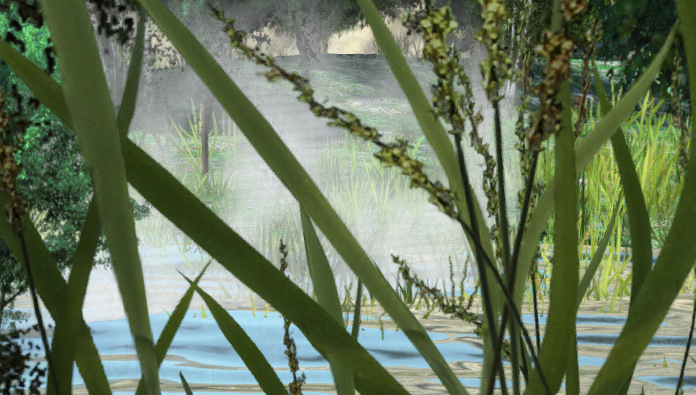}}

\end{tabular}
\caption{Comparison of Phenomenological Transparency variants to an offline A-Buffer. 
Compared to the state of the art Moments, at 3rd order Wavelets are an alternative approach 
with similar quality that is more than 4x bandwidth efficient and significantly faster when the 
image is dominated by opaque surfaces.
Rows 1-5: Quality of Wavelet and Moments are similar to each other and the A-buffer reference, and superior to previous methods for smoke and glass quality.
Rows 6-7: Wavelets produce less of the dark ringing in the foliage than previous methods at the transition from partly to fully covered.
Rows 8-9: Wavelets produce less of the dark ringing in the edges of the foliage. Otherwise the quality is comparable to Moments in a scene which combines foliage, particles and water.
}
\label{fig:comparison}
\end{figure*}

\begin{figure*}[tb]
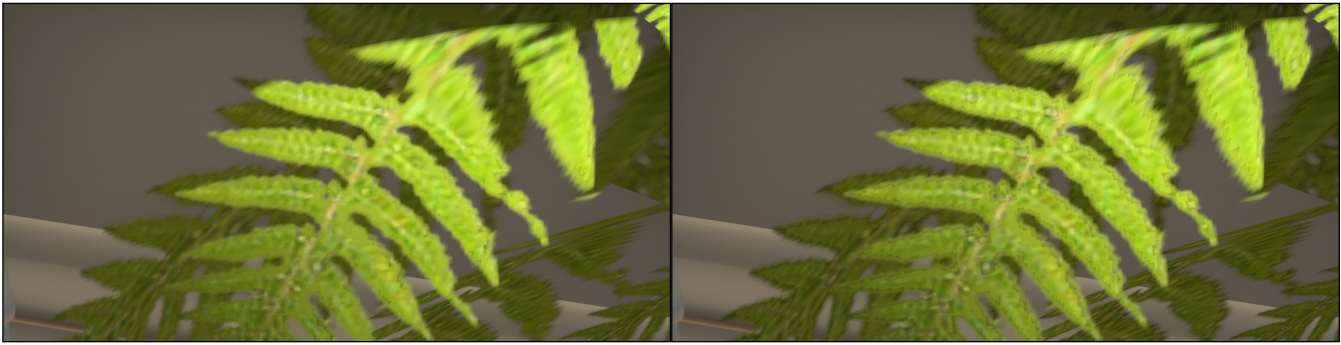

    \centering
    \subfloat[Wavelet]{\fbox{\includegraphics[width=0.5\textwidth,clip,trim=0 0 0 1cm]{results/leaves1/wavelet_r3_in.jpg}}}
    \subfloat[Moments]{\fbox{\includegraphics[width=0.5\textwidth,clip,trim=0 0 0 1cm]{results/leaves1/moments_o6_in.jpg}}}
    \caption{Zoomed in comparison of Wavelets and Moments with tuned bias. Notice that Wavelets have significantly less ringing on the edges. If the bias in Moments is tuned to handle transparency as well, then the ringing worsens as can be seen
    in Figure~\ref{fig:comparison} Row 9.}
    \label{fig:leaves}
\end{figure*}

\begin{figure*}[tb]
\centering
\subfloat[No refraction]{\fbox{\includegraphics[width=0.33\textwidth,clip,trim=0 0 0 1cm]{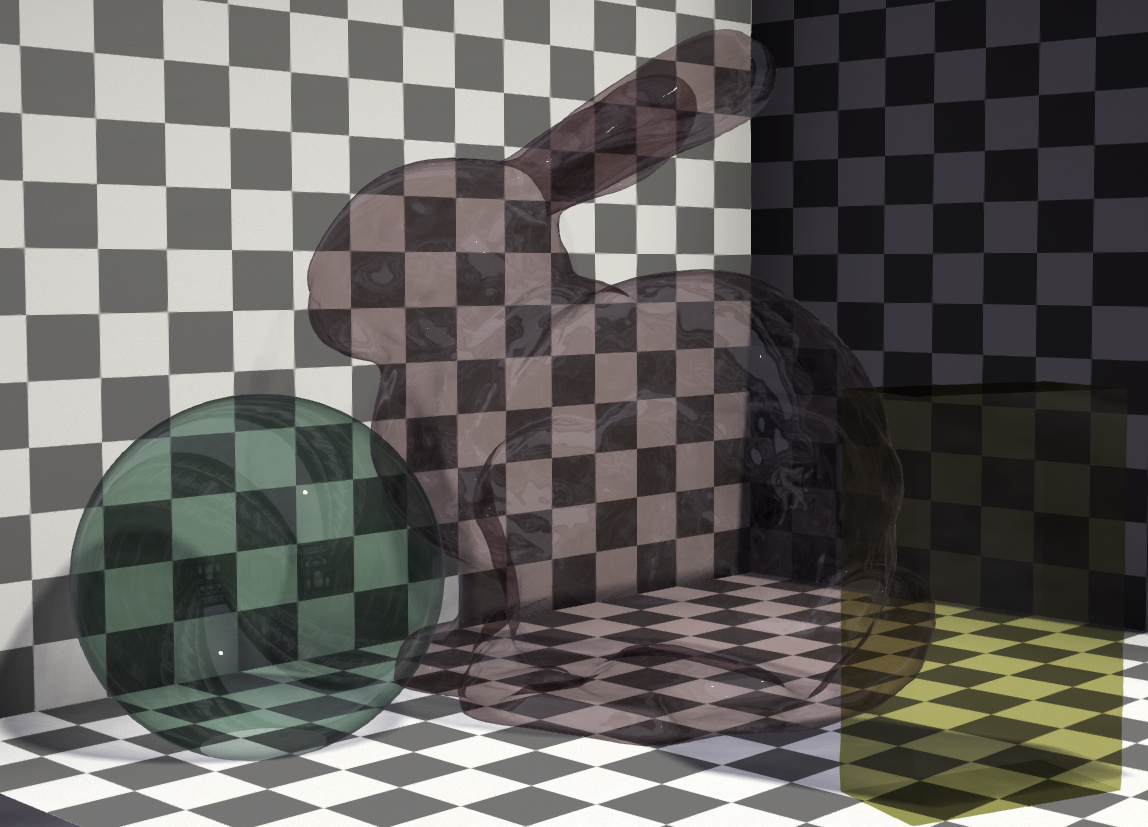}}}
\subfloat[Refraction]{\fbox{\includegraphics[width=0.33\textwidth,clip,trim=0 0 0 1cm]{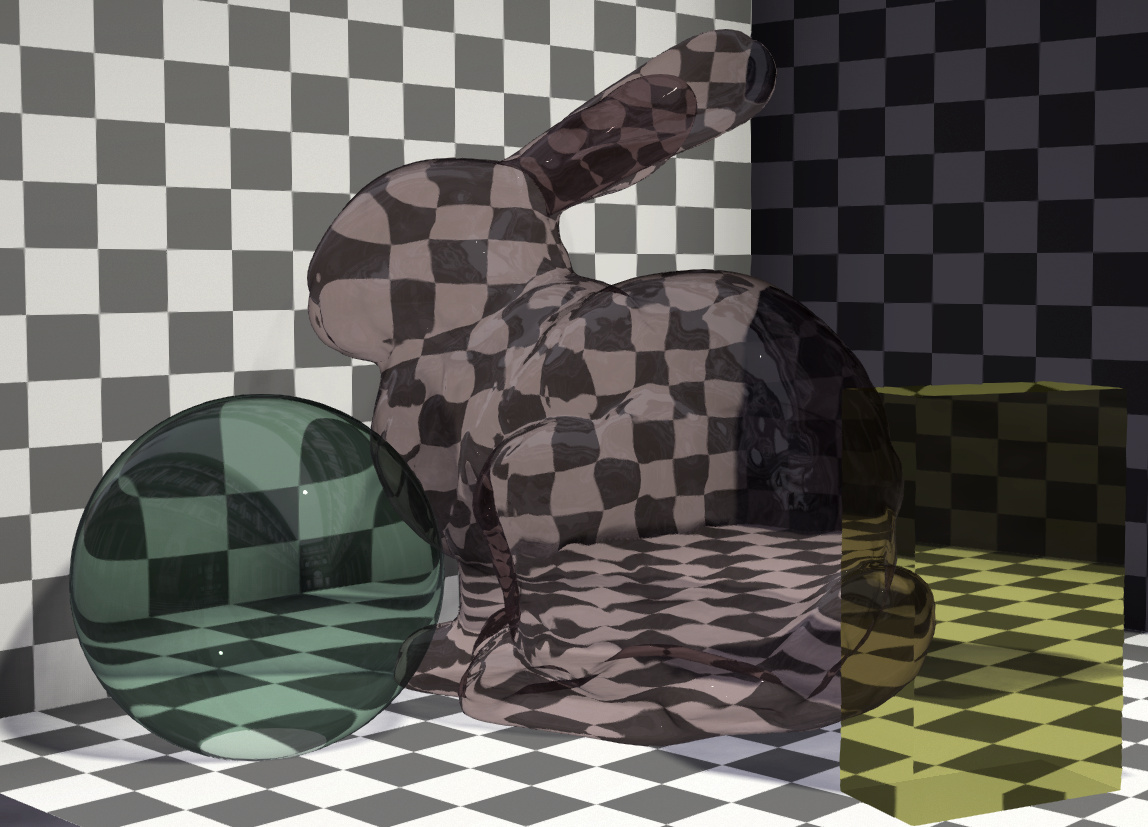}}}
\subfloat[Refraction + new chromatic aberration]{\fbox{\includegraphics[width=0.33\textwidth,clip,trim=0 0 0 1cm]{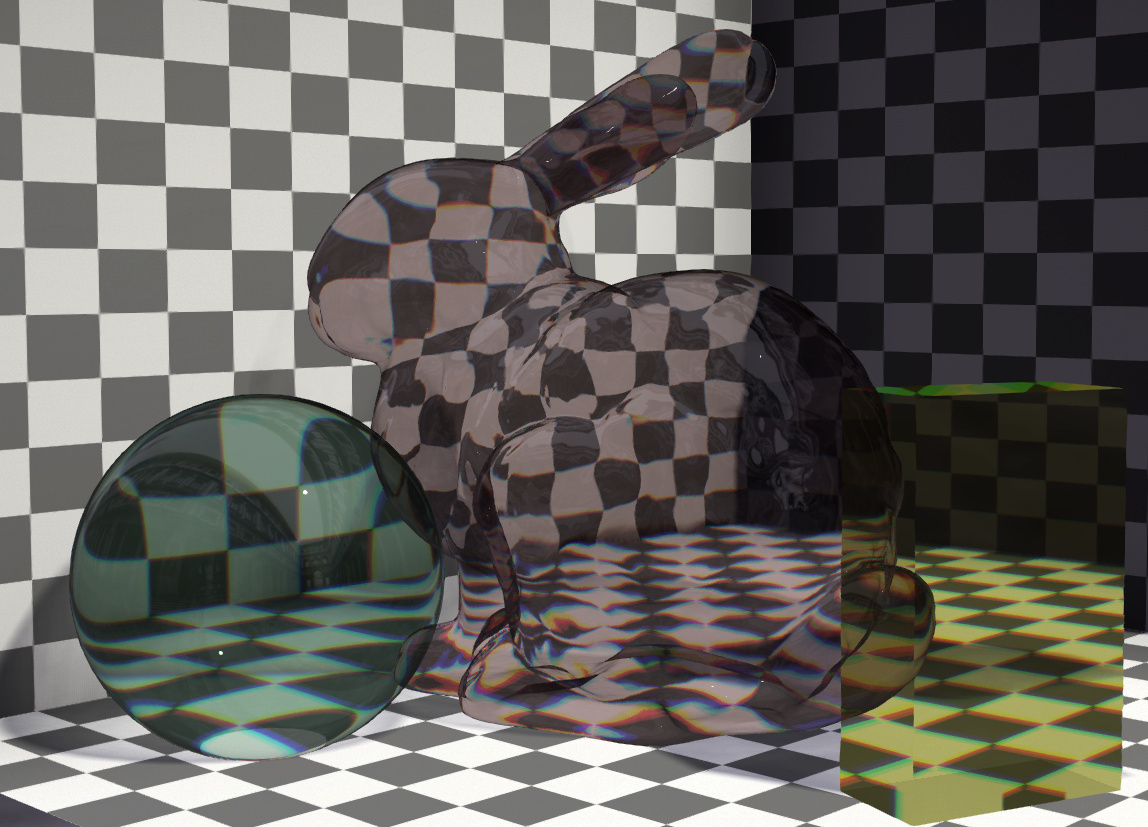}}}
\caption{Tinted glass objects under increasingly realistic rendering.}
\label{fig:chromatic}
\end{figure*}

\begin{figure*}
\centering
\fbox{\includegraphics[width=0.5\textwidth,clip,trim=9cm 0 9cm 0]{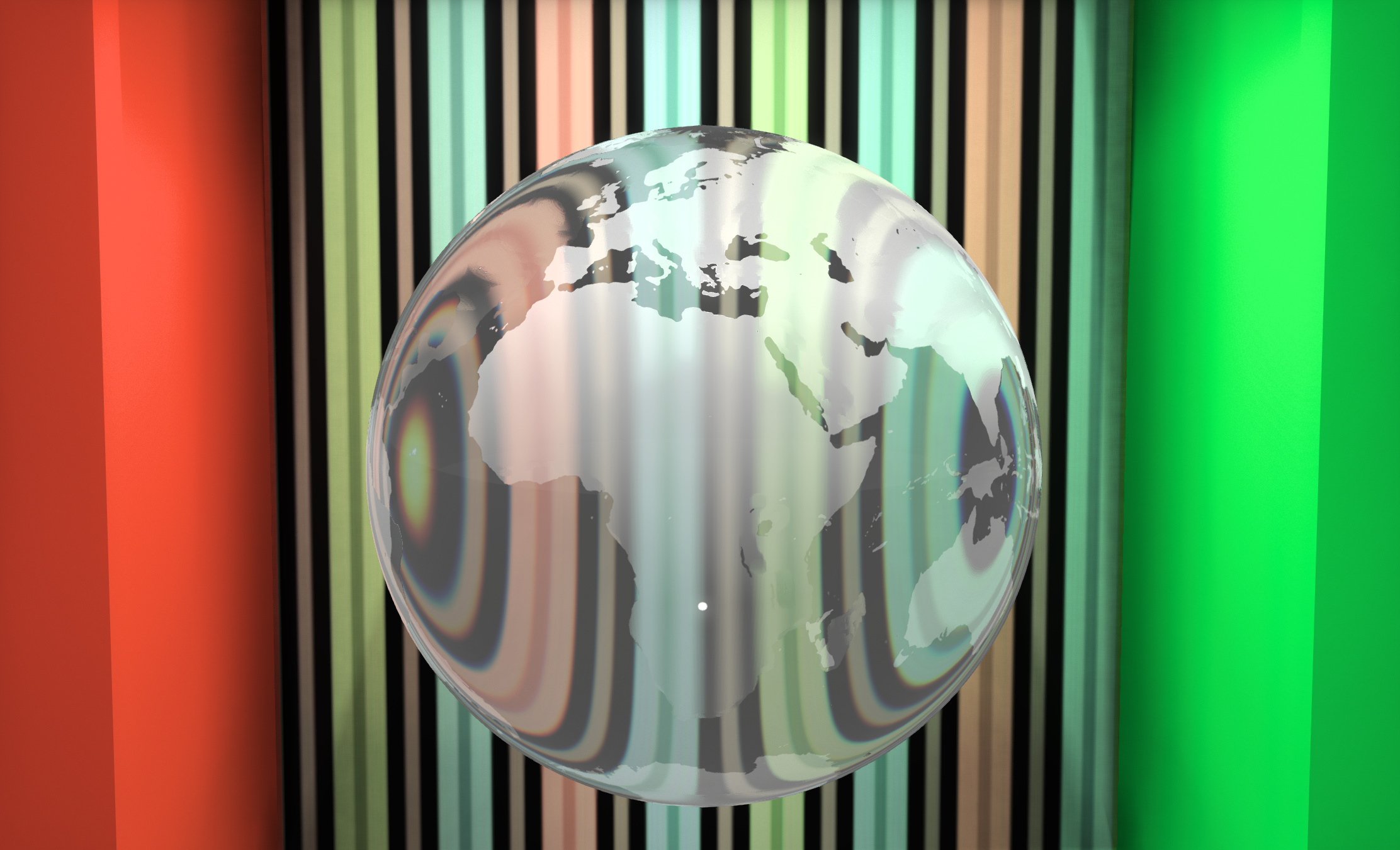}}
\caption{Refraction, dispersion, and chromatic aberration phenomena.}
\label{fig:refract-globe}
\end{figure*}

\subsection{Quality Comparison}
The grid in Figure~\ref{fig:comparison} compares three recent OIT strategies to the new Wavelet Transparency and to a non-realtime A-buffer as a reference. We implemented each of these methods as a replacement for the visibility strategy for Phenomenological Transparency but retained the full suite of other effects that algorithm produces. In the first two rows, Moment8 and Wave4 are used because Moment6 underflowed on this scene. All other rows use Moment6 and Wave3. We use single-precision Moment instead
of quantized to maximize its image quality (for performance comparisons we used 16-bit quantization). MLAB uses four layers in all rows.

Row 1 shows an opaque sphere in a particle system against an orange background. Row 2 repeats this scene with a transparent sphere. In each case MLAB4 underperforms on coverage and WBOIT fails badly, revealing too much of the sphere. Moment and wavelet transparency each give similar results, which are close to the ideal A-buffer result.

Rows 3-5 are different views of the Bistro scene with many layers of drinking glasses. Where the number of layers is small, all algorithms are similar. When the number of layers becomes high it exceeds the MLAB's buffer, as seen at the reflection on the first glass just left of center in row 5. In these cases WBOIT fails more gradually but presents an ambiguous result, where the ordering of glasses in depth becomes less obvious. Moment is much better, but oversmooths the visibility function and tends to excessively dim the highlights from glasses beyond the first layer. Wavelets are indistinguishable from the A-buffer, as the step functions due to glass are the ideal case for the Haar basis.

Rows 6-7 show alpha-masked leaves from two viewpoints. Examining the coverage transition at the edges of leaves, especially where those edges overlap in depth, we observe that MLAB underestimates coverage, WBOIT and moment have some noise and discontinuity, and Wavelet most closely (but not perfectly) matches the reference A-buffer.

We conclude that across all results, Wavelet is the most robust because it is consistently among the best approximations, even though in specific cases others may also perform well.

\subsection{Chromatic Aberration}

Figure~\ref{fig:chromatic} shows a simplified scene to highlight the new chromatic aberration approximation (all other figures in this paper used it with refraction as well). On the left is the scene without refraction simulation, in the center is the method of McGuire and Mara~\cite{McGuire2017Transparency}, and in the right is the new chromatic aberration, which is most noticable at high-contrast edges. Compare this to real chromatic aberration photographed in Figure~\ref{fig:aberration}. Despite a tenuous relationship to the physics of the real effect, we suggest that the simple simulation in (c) yields a more convincing image of the refraction phenomenon than without it in (b).

Figure~\ref{fig:refract-globe} is inspired by a famous path traced figure by Walter et al.~\cite{Walter:ea:2007}.
At low cost in a rasterization pixel shader pass, our method demonstrates the phenomena of refraction, diffusion, and chromatic aberration of a frosted glass globe. The refraction pattern differs because they modeled a solid ball and ours is a hollow sphere with thick walls. 

\subsection{Performance}

Table~\ref{tbl:perf} shows the time in milliseconds for steps 2 and 3: building the visibility approximation and then evaluating it and shading, for the $1920\times1080$ rendering of the view of drinking glasses on the Bistro bar in row 5 of Figure~\ref{fig:comparison} on GeForce RTX 3090. The composite time for step 4 is the same for all algorithms: 0.1ms. Performance of Moment Transparency is done here in 16-bit quantized precision mode.

\begin{table}[h]
    \begin{center}
        \begin{tabular}{r|r|r|r|r|r}
            MLAB4 & WBOIT &  Moment6 & Moment8 &  Wavelet3 & Wavelet4 \\
            \hline
            1.8 & 0.46 &  2.04 &  2.1 &   1.8 & 2.05\\
        \end{tabular}
    \end{center}
    \caption{Visibility and shading time in milliseconds}
    \label{tbl:perf}
\end{table}

\section{Discussion}
Previous methods such as Fourier opacity maps read and write \textit{all} coefficients in the visibility approximation at a pixel at every step. Due to the hierarchical compact support, wavelets require only a logarithmic number of basis evaluations in the extent of the represented function. This gives their superior asymptotic bandwidth bound. Because constant factors differ across bases, we chose the minimal 1st-order Daubechies, a.k.a. Haar, basis. For the asymptotically optimally higher-order Daubechies basis~\cite{Daubechies1992Wavelet}, the constant factor impractically large for real-time visibility function approximation today.

Furthermore, the wavelet coefficients are calculated by adding same signed and non dimminishing values. As a result, the coefficients themselves are also non possitive for all wavelet functions. Overall the algorithm exhibits superior numerical robustness compared to other methods.


The decimated wavelet transform we employed lacks shift invariance. Therefore, the visibility approximation of a fixed scene can change under camera motion along the view vector. We propose future work exploring the use of \textit{complex} wavelets~\cite{Selesnick2005CWT}, which are nearly shift invariant, although more difficult to optimize. \textit{Stationary} wavelets are also shift invariant, although require too many coefficients to be viable for this application.


Fourier Opacity Maps, Moment Transparency, and Wavelet OIT share the limitation that they must make two passes over transparent geometry per frame. To avoid this for Wavelets we propose computing the coefficients in the same pass as shading. This of course means that the current coefficients are not available to the shading. So, when reading the coefficients to reconstruct the visibility function, the shading routine uses the transformations from the \textit{previous frame} to read at the projected positions from the previous wavelet buffer. A guard band can handle clipping at the edge of the screen, but efficiently handling occlusions with opaque objects on screen is one open problem for this optimization.

%
%
%

\clearpage

\bibliographystyle{eg-alpha-doi}  
\bibliography{c:/Users/maizenstein/Documents/wavelet-transparency/paper/arxiv_submission/woit}

\end{document}